\documentclass[twocolumn]{aastex631}
\usepackage{amsmath,booktabs,multirow}
\usepackage{tikz}
\usepackage{tikz-3dplot}
\usetikzlibrary{arrows.meta}
\usepackage{float}
\usepackage{placeins}
\usepackage{longtable}

\newcommand{\qpefit}{\href{https://github.com/joheenc/QPE-FIT/tree/main}{\texttt{QPE-FIT}}}

\shorttitle{QPE timing with Bayesian Inference}
\shortauthors{Chakraborty, Drummond \textit{et al.}}

\begin{document}

\title{ Prospects for EMRI/MBH parameter estimation using Quasi-Periodic Eruption timings: short-timescale analysis}

\author[0000-0002-0568-6000]{Joheen Chakraborty}
\thanks{Both authors contributed equally to this work}
\affiliation{Department of Physics \& Kavli Institute for Astrophysics and Space Research, Massachusetts Institute of Technology, Cambridge, MA}

\author[0000-0002-8435-9955]{Lisa V.\ Drummond}
\thanks{Both authors contributed equally to this work}
\affiliation{Department of Physics and TAPIR, California Institute of Technology, Pasadena, CA}

\author[0000-0001-7889-6810]{Matteo Bonetti}
\affiliation{Dipartimento di Fisica ``G. Occhialini'', Universit\`a degli Studi di Milano-Bicocca, Piazza della Scienza 3, I-20126 Milano, Italy}
\author[0000-0002-8400-0969]{Alessia Franchini}
\affiliation{Dipartimento di Fisica ``G. Occhialini'', Universit\`a degli Studi di Milano-Bicocca, Piazza della Scienza 3, I-20126 Milano, Italy}

\author[0009-0004-5838-1886]{Shubham Kejriwal}
\affiliation{Department of Physics, National University of Singapore, Singapore 117551}

\author[0000-0003-0707-4531]{Giovanni Miniutti}
\affiliation{Centro de Astrobiolog\'ia (CAB), CSIC-INTA, Camino Bajo del Castillo s/n, 28692 Villanueva de la Ca\~nada, Madrid, Spain}

\author[0000-0003-4054-7978]{Riccardo Arcodia}
\thanks{NHFP Einstein Fellow}
\affiliation{Department of Physics \& Kavli Institute for Astrophysics and Space Research, Massachusetts Institute of Technology, Cambridge, MA}

\author[0000-0001-6211-1388]{Scott A. Hughes}
\affiliation{Department of Physics \& Kavli Institute for Astrophysics and Space Research, Massachusetts Institute of Technology, Cambridge, MA}

\author[0000-0003-0743-6491]{Francisco Duque}
\affiliation{Max Planck Institute for Gravitational Physics (Albert Einstein Institute) Am Mühlenberg 1, D-14476 Potsdam, Germany}

\author[0000-0003-0172-0854]{Erin Kara}
\affiliation{Department of Physics \& Kavli Institute for Astrophysics and Space Research, Massachusetts Institute of Technology, Cambridge, MA}

\author[0000-0003-4961-1606]{Alberto Sesana}
\affiliation{Dipartimento di Fisica ``G. Occhialini'', Universit\`a degli Studi di Milano-Bicocca, Piazza della Scienza 3, I-20126 Milano, Italy}

\author[0000-0002-1329-658X]{Margherita Giustini}
\affiliation{Centro de Astrobiolog\'ia (CAB), CSIC-INTA, Camino Bajo del Castillo s/n, 28692 Villanueva de la Ca\~nada, Madrid, Spain}

\author{Amedeo Motta}
\affiliation{Dipartimento di Fisica ``G. Occhialini'', Universit\`a degli Studi di Milano-Bicocca, Piazza della Scienza 3, I-20126 Milano, Italy}

\author[0000-0002-7226-836X]{Kevin Burdge}
\affiliation{Department of Physics \& Kavli Institute for Astrophysics and Space Research, Massachusetts Institute of Technology, Cambridge, MA}

\begin{abstract}

Quasi-Periodic Eruptions (QPEs) are luminous, recurring X-ray outbursts from galactic nuclei, with timescales of hours to days. While their origin remains uncertain, leading models invoke accretion disk instabilities or the interaction of a massive black hole (MBH) with a lower-mass secondary in an extreme mass ratio inspiral (EMRI). EMRI scenarios offer a robust framework for interpreting QPEs by characterizing observational signatures associated with the secondary's orbital dynamics. This, in turn, enables extraction of the MBH/EMRI physical properties and provides a means to test the EMRI scenario, distinguishing models and addressing the question: \textit{what can QPE timings teach us about massive black holes and EMRIs?} In this study, we employ analytic expressions for Kerr geodesics to efficiently resolve the trajectory of the secondary object and perform GPU-accelerated Bayesian inference to assess the information content of QPE timings. Using our inference framework, referred to as \href{https://github.com/joheenc/QPE-FIT/tree/main}{\texttt{QPE-FIT} (Fast Inference with Timing)}, we explore QPE timing constraints on astrophysical parameters, such as EMRI orbital parameters and MBH mass/spin. We find that mild-eccentricity EMRIs ($e\sim0.1-0.3$) can constrain MBH mass and EMRI semimajor axis/eccentricity to the 10\% level within tens of orbital periods, while MBH spin is unconstrained for the explored semimajor axes $\geq 100R_g$ and monitoring baselines $\mathcal{O}(10-100\rm)$ orbits. Introducing a misaligned precessing disk generally degrades inference of EMRI orbital parameters, but can constrain disk precession properties within 10-50\%. This work both highlights the prospect of QPE observations as dynamical probes of galactic nuclei and outlines the challenge of doing so in the multimodal parameter space of EMRI-disk collisions.

\end{abstract}

\section{Introduction} \label{sec:intro}
Quasi-periodic eruptions (QPEs) are a recently identified class of high-amplitude  soft X-ray flares recurring on timescales of hours to days, typically observed from galactic nuclei hosting low-mass massive black holes (MBHs). The known QPE sample has grown to around eleven sources \citep{Miniutti2019, Giustini2020, Arcodia2021, Arcodia2024a, Chakraborty2021, Chakraborty2025a, Quintin2023, Nicholl2024, HernandezGarcia2025}, enabled primarily by high-cadence X-ray monitoring from \textit{XMM-Newton} and \textit{NICER}. These systems exhibit repeating outbursts with recurrence times that are generally stable over many cycles. The flare energetics and recurrence behavior point to a compact, coherent engine that is both localized (operating in the immediate environment of the MBH) and temporally stable across multiple eruption episodes. However, the physical origin of QPEs remains under active investigation, with three classes of theoretical frameworks emerging in recent years.

One class of models attributes QPEs to intrinsic variability in the accretion flow. Proposed mechanisms include radiation-pressure instabilities and the relativistic precession of a warped or geometrically thick inner disk around a spinning MBH \citep{Raj2021,Pan2021_2,Pan2022,Pan2023,Kaur2023,Sniegowska2020,Sniegowska2023,Middleton2025}. These scenarios can naturally produce variable accretion flows, but face challenges explaining the QPE recurrence times, spectral evolution, and long viscous timescales compared to the observed eruption durations. Continued investigation is nevertheless warranted to further explore QPE observables using disk instability models.

A second class of interpretations focuses on mass transfer in extreme mass-ratio inspirals (EMRIs). EMRIs are gravitationally bound binaries which consist of a stellar-mass object (e.g. a star or stellar-remnant black hole) orbiting a supermassive black hole (MBH), with the inspiral driven predominantly by gravitational-wave emission. We refer to the stellar-mass object as the ``secondary". In this class of models, the eruptions are powered by episodic accretion from the secondary onto the MBH \citep{King2020, King2022, King2023, Metzger2022, Zhao2022, Chen2022, Krolik2022, Linial2023a, Yang2025, Lau2025}. These models naturally yield periodicity tied to the orbital motion, but still require rapid accretion compared to the viscous timescale, raising questions about the feasibility of this scenario. Additionally, these models often produce clocks that are \textit{too} stable, in contrast to the slight quasi-periodicity and frequency wandering observed in some QPE sources.

A third possibility, which is gaining increasing attention, proposes that QPEs are powered by repeated collisions between a compact secondary and the accretion disk of the MBH. In this picture, the secondary plunges through the disk typically twice per orbit, generating shocks, outflows, or thermal transients that give rise to the observed X-ray flares \citep{Dai2010, Xian2021, Sukova2021, Lu2023, Franchini2023, Tagawa2023, Linial2023b, Linial2024b, Zhou2024a,Zhou2024c,Zhou2024b,Zhou2025,Lam2025,Yao2025a,Yao2025b,Huang2025}. This model combines the advantages of a stable orbital clock from EMRIs with a physically motivated and localized energy release mechanism. Crucially, the eruption timings correspond to the approximately geodesic motion of the secondary and are largely independent of viscous timescales, allowing for a robust link between the flare cadence and underlying orbital dynamics. We caution that the EMRI collisional model is far from confirmed as the true physical origin of QPEs---see e.g. \cite{Mummery2025}, \cite{GuoShen2025}, and \cite{Guolo2025a} for further discussion on the shortcomings of these models in adequately producing the energetics, temperatures, and scaling relations predicted by the simple-case pictures.  Additional tests across a variety of observables, e.g.\ timing residuals \citep{Chakraborty2024,Arcodia2024b,Pasham24b,Giustini2024,Miniutti2025,Zhou2024b} and spectral/SED analysis \citep{Wevers2025,Guolo2025a,Chakraborty2025b,GuoShen2025} are well-motivated, and are indeed an increasing focus in the literature.

In the EMRI model scenario, some studies also propose that QPEs are formed when the orbiter begins interacting with the compact accretion disks formed in the aftermath of tidal disruption events (TDEs) \citep{Linial2023b,Franchini2023}. In this EMRI+TDE framework, the secondary predominantly inspirals via GW emission and two-body scattering, until it reaches $\mathcal{O}(100 R_g)$, where it intersects the TDE disk. This scenario is especially compelling for QPEs where a TDE was explicitly observed several years prior to the onset of the eruptions \citep{Quintin2023,Nicholl2024,Bykov2025,Chakraborty2025a}, and is bolstered by mounting evidence for long-lived precursor flares in other systems without a definitive TDE association \citep{Chakraborty2021,Miniutti2023a,Arcodia2024a,Guolo2025b}.

QPEs may represent the first electromagnetic detection of EMRIs, one of the key anticipated gravitational-wave (GW) sources targeted by the upcoming space-based millihertz observatories such as \textit{LISA} \citep{AmaroSeoane2023} and \textit{TianQin} \citep{Luo2016}. EMRIs offer extremely precise tests of strong-field gravity and detailed probes of massive black hole environments, motivating their importance in the \textit{LISA} science case \citep{Barack2004}. It has been proposed that ``wet" EMRIs (those formed in the presence of an accretion disk) may constitute a common formation channel for \textit{LISA} sources \citep{Pan2021,2024MNRAS.528.4958W}, as their inward migration rate may be significantly enhanced via dissipative interactions with the accretion flow. EMRIs formed in these matter-rich environments can carry imprints of the accretion disk in their gravitational wave signals \citep{Derdzinski2021,Speri2023,Copparoni2025,Duque2025}, and may also produce quasi-periodic electromagnetic signatures during their inspiral.  

Wet EMRIs have therefore been proposed as a possible source of periodic soft X-ray variability at millihertz frequencies, including quasi-periodic oscillations (QPOs) and QPEs, and may enable multi-messenger observations of individual systems \citep{Kejriwal2024,Suzuguchi:2025lyx}. While the population of thus-far known QPEs is at periods too long to be detectable by mHz observatories \citep{Kara2025}, a subset of \textit{LISA}-detectable EMRIs may emit X-rays in the same frequency band as observed QPOs \citep{Gierlinski2008,Masterson2025}, making them possible multi-messenger candidates. There is therefore growing potential for current and future X-ray observatories to identify EMRIs in advance of \textit{LISA}’s launch.

In principle, the EMRI collisional model encodes information about the orbiter-disk system via the QPE timings (assuming a constant delay between collision and peak light). In this work, we address the question: \textit{how much information about the central MBH and the orbit of the secondary can be extracted from QPE timing data alone?} We assess how different observing strategies, as well as different astrophysical parameters, impact inference. Using an efficient Post-Newtonian expansion for describing Kerr geodesics and including the impact of disk precession and geometric/Shapiro delays, we develop a computationally optimized, GPU-accelerated framework for simulating the timing of eruptions and assessing the prospects for astrophysical inference from existing and future observations. We make our QPE timing calculation code, dubbed \qpefit\ (Fast Inference with Timing), publicly available on GitHub\footnote{\href{https://github.com/joheenc/QPE-FIT/tree/main}{https://github.com/joheenc/QPE-FIT}}, to  support future observational data fitting and serve as a resource for the community. \qpefit\ will undergo active development, and the most up-to-date features will be described at the provided link.

In Section~\ref{sec:model} we describe the features of our model, including the EMRI orbital trajectory calculation, disk prescription, QPE timing method, and likelihood function for Bayesian inference. In Section~\ref{sec:results} we show the results of applying our inference algorithm across a large grid of simulated QPE timings to assess how EMRI/disk properties affect parameter inference efficacy, and fit the timing data of eRO-QPE1 to show that the precessing disk model is able to describe the erratic timing behaviors observed in some QPEs. We discuss the possibility of further astrophysical inference from QPE timings in Section~\ref{sec:discussion}, and make concluding remarks in Section~\ref{sec:conclusion}.

\section{Methods} \label{sec:model}
In this paper, we design our timing model to achieve the following objectives:
\begin{enumerate}
    \item Develop an efficient framework for QPE source analysis that can be applied broadly and evaluated quickly. This is accomplished by (i) using a Post-Newtonian (PN) model for Kerr geodesics, which is computationally efficient and described in detail in Sec.\ \ref{sec:trajectory}; and (ii) using GPU-acceleration. Evaluation speed is crucial to allow Bayesian parameter inference, which typically entails $>$millions of model evaluations. We note that the PN formalism provides limited accuracy for low orbital separations ($\lesssim 50R_g$) and large eccentricities ($e\gtrsim 0.5$); we explore its limitations in Appendix~\ref{subsec:pn_vs_kerr}.
    \item Create a flexible model that accounts for diverse QPE timing behaviors, including sources with irregular timing patterns. To achieve this, we model flare timings as interactions between the secondary object and a rigidly precessing accretion disk. The disk precession model is discussed in \ref{sec:disk}.
    \item Coherently resolve flare timings while accounting for observational gaps. This is achieved by constructing the flare timing likelihood to match observed flares to model timings, enabling accurate parameter estimation even when large sections of the dataset are omitted. We use this framework to evaluate different observing strategies and assess their relative abilities to recover EMRI and disk parameters. We note that in this paper, we only model short observational gaps/baselines, of order tens to hundreds of consecutive QPEs, and defer long-term monitoring including secular evolution effects to future work.
\end{enumerate}

This approach ensures our timing model is both adaptable and practical for analyzing QPE sources with varied timing behaviors and observational challenges. We build heavily on the existing work of \cite{Franchini2023}, with the key addition that our computational accelerations allow for full Bayesian inference. To illustrate the efficiency gains, we benchmark the speed of \textit{one evaluation} of the log-likelihood function with and without a GPU. We used an NVIDIA A100 GPU and an AMD EPYC 7763 256-core processor operating with a 2652.814 MHz clock speed; the resulting wall-clock times of the computation steps are shown in Table~\ref{tab:benchmark}. We performed two separate benchmarks using 100 and 1000 parallel walkers. As is typical for GPU-accelerated programs, the best improvements are seen for higher parallelism, with a speed-up of $26880\times$ comparing one GPU to one CPU for the 1000-walker case. These performance gains enable rapid posterior exploration using ensemble samplers, underscoring the practical utility of \qpefit\ for future population-level studies.

\begin{table*}[ht]
\centering
\caption{Typical speed-up from GPU-accelerating \textit{one evaluation} of the likelihood function. We used an NVIDIA A100 GPU, compared with an AMD EPYC 7763 256-core CPU operating with a 2652.814 MHz clock speed.}
\label{tab:benchmark}
\begin{tabular}{ccccc|cccc}
\toprule
 & \multicolumn{4}{c|}{100 walkers} & \multicolumn{4}{c}{1000 walkers} \\
\cmidrule(lr){2-9}
 & CPU & GPU & Gain & Gain & CPU & GPU & Gain & Gain \\
Computing step & (sec) & (sec) & (wall time) & (per core) & (sec) & (sec) & (wall time) & (per core) \\
\midrule
EMRI trajectory & 0.232 & 0.0119 & $19.5\times$ & $4992\times$ & 2.37 & 0.0119 & $199\times$ & $50944\times$ \\
Disk crossings & 0.0257 & 0.00199 & $12.9\times$ & $3302\times$ & 0.352 & 0.0136 & $25.9\times$ & $6630\times$ \\
Timing residuals & 0.00395 & 0.00142 & $2.78\times$ & $712\times$ & 0.0439 & 0.00452 & $9.71\times$ & $2486\times$ \\
\textbf{Total log-likelihood} & 0.2622 & 0.0154 & $17.0\times$ & $\mathbf{4352\times}$ & 2.77 & 0.0263 & $105\times$ & $\mathbf{26880\times}$\\
\bottomrule
\end{tabular}
\end{table*}

\subsection{Trajectory of the orbiting body} \label{sec:trajectory}
The trajectory of the secondary object in the QPE timing model is governed by bound geodesics in Kerr spacetime. The Kerr metric, describing a rotating black hole with spin parameter $\chi_\bullet$, is expressed in Boyer-Lindquist coordinates $\{t,r,\theta,\phi\}$ as follows:
\begin{align}
\label{eq:kerr}
d {s}^2 &= 
-\left(1 - \frac{2r}{\Sigma}\right) d{t}^2 
+ \frac{\Sigma}{\Delta} d{r}^2
+ \frac{\Sigma}{1-z^2} d{z}^2\nonumber
\\
&+ \frac{1-z^2}{\Sigma} \left(2\chi_\bullet^2 r (1-z^2)+(\chi_\bullet^2+r^2)\Sigma\right) d\phi^2
\nonumber\\&- \frac{4\chi_\bullet r(1-z^2)}{\Sigma} d{t}d\phi,
\end{align}
where $z=\cos\theta$, $\Delta= r(r-2)+\chi_\bullet^2$  and $\Sigma = r^2 + \chi_\bullet^2 z^2$. In the above equations, the black hole's mass is normalized to unity ($M_\bullet=1$).\footnote{Throughout this work, $\chi_\bullet$ denotes the dimensionless spin of the massive black hole, with the mass normalized to $M_\bullet = 1$. We restore physical units, including the mass scale, when computing observables such as flare timings. For reference, $GM_\bullet/c^3 = 49.2\,{\rm sec}(M_\bullet/10^7M_\odot)$.} 

The Kerr metric admits four constants of motion that facilitate analytic solutions for geodesic motion. The first conserved quantity, $u_\mu u^\mu$, arises from the definition of proper time, $\tau$. The specific energy $\mathcal{E}$ and axial angular momentum $\mathcal{L}_z$ are conserved due to the spacetime's Killing vector fields, while the existence of a Killing tensor leads to the Carter constant $\mathcal{Q}$.  In the Schwarzschild limit, $\mathcal{Q}$ is simply related to the total angular momentum of an orbit: $\mathcal{Q}(\chi_\bullet = 0) = |\boldsymbol{\mathcal{L}}|^2 - \mathcal{L}_z^2$.  For $a_{\bullet} \ne 0$, this correspondence does not hold precisely, but provides useful intuition about the meaning of $\mathcal{Q}$, especially for orbits with wide separation.

The radial and polar motion of the geodesic can be described analytically using elliptic functions \cite{Schmidt2002,Fujita2009,vandeMeent2020}: 
\begin{align}
r(\lambda) &= \frac{r_3 (r_1 - r_2) \, \text{sn}^2\left(\frac{\mathsf{K}(k_r) q_r}{\pi} \, \Big| \, k_r\right) - r_2 (r_1 - r_3)}{(r_1 - r_2) \, \text{sn}^2\left(\frac{\mathsf{K}(k_r) q_r}{\pi} \, \Big| \, k_r\right) - (r_1 - r_3)}, \\
z(\lambda) &= z_{1} \, \text{sn}\left(\mathsf{K}(k_z) \frac{2q_z}{\pi} \, \Big| \, k_z\right).
\end{align}
Here, $r_1$, $r_2$, and $z_1$ are the radial and polar turning points, while $k_r$ and $k_z$ are given by
\begin{align}
k_r &=\frac{(r_1-r_2)(r_3-r_4)}{(r_1-r_3)(r_2-r_4)},
\\
k_z &= \chi_\bullet^2 (1-\mathcal{E}^2)\frac{z_1^2}{z_2^2}.
\end{align}
The roots $r_3$, $r_4$ and $z_2$ are determined by the radial and polar potentials, as described by Eqs.\ (13)--(14) in \cite{vandeMeent2020}. The motion is parameterized in terms of the Mino time parameter $\lambda=\int d\tau/\Sigma$ \citep{Mino2003} which decouples the radial and polar motion, thus simplifying the geodesic equations.

The functions $\mathsf{K}(k)$ and $\text{sn}(u|k)$ represent the complete elliptic integral of the first kind and the Jacobi elliptic sine function, respectively. The phases $q_r$ and $q_z$ evolve linearly with Mino time as: 
\begin{align}
q_r &= \Upsilon_r \lambda + q_{r, 0}, \quad q_z = \Upsilon_z \lambda + q_{z, 0},
\end{align}
where $ \Upsilon_r $ and $ \Upsilon_z $ are the Mino-time frequencies, and $ q_{r, 0} $ and $ q_{z, 0} $ are initial phases. Each bound Kerr geodesic can be uniquely mapped to a set of quasi-Keplerian orbital parameters that offer a more intuitive description of the motion. These include the black hole spin $\chi_\bullet$, semi-latus rectum $p$, eccentricity $e$, inclination angle $i$, and three phase constants ($q_{r,0},q_{z,0},q_{\phi,0}$) that specify the initial conditions. This seven-dimensional parameterization fully characterizes the geodesic and forms the basis of our trajectory model.

While the orbital inclination $i$ provides an intuitive measure of tilt relative to the black hole spin, it is important to note that generic bound geodesics in Kerr spacetime do not lie in a fixed plane. Instead, the motion fills a two-dimensional torus due to independent radial and polar oscillations. The inclination angle $i$ used here refers to a conserved parameter derived from the orbital constants of motion via $\cos i =\mathcal{L}_z/\sqrt{\mathcal{L}_z^2+Q}$ and should not be interpreted as the geometric angle between a unique orbital plane and the black hole spin axis.

The first derivatives of $ r $ and $ z $ with respect to Mino time are determined by the potentials $ R(r) $ and $ Z(z) $:
\begin{align}
\left(\frac{dr}{d\lambda}\right)^2 &\equiv R(r) \nonumber= [(r^2 + \chi_\bullet^2)\mathcal{E} - \chi_\bullet\mathcal{L}_z]^2 \\&- \Delta[r^2 + (\mathcal{L}_z - \chi_\bullet\mathcal{E})^2 + \mathcal{Q}], \\
\left(\frac{dz}{d\lambda}\right)^2 &\equiv Z(z) = \mathcal{Q}(1 - z^2) \nonumber \\& - \left[(1 - z^2)(1 - \mathcal{E}^2)\chi_\bullet^2 + \mathcal{L}_z^2\right]z^2.
\end{align}
The geodesic motion also includes azimuthal and temporal components, which are expressed as:
\begin{align}
\frac{d\phi}{d\lambda} &\equiv \Phi(r,z) = \frac{\chi_\bullet}{\Delta} \big[\mathcal{E}(r^2 + a^2) - \chi_\bullet\mathcal{L}_z\big] \nonumber\\&+ \frac{\mathcal{L}_z}{1 - z^2} - \chi_\bullet\mathcal{E}, \\
\frac{dt}{d\lambda} &\equiv T(r,z) = \frac{r^2 + \chi_\bullet^2}{\Delta} \big[\mathcal{E}(r^2 + \chi_\bullet^2) - \chi_\bullet\mathcal{L}_z\big] \nonumber\\&- a^2 \mathcal{E}(1 - z^2) + \chi_\bullet\mathcal{L}_z.
\end{align}
The temporal component relates Mino time $\lambda$ to the coordinate time $t$.

While the exact Kerr geodesic equations presented above enable high-precision modeling of the secondary’s trajectory, their numerical implementation can be computationally intensive, particularly when incorporated into a Bayesian inference pipeline which requires repeated trajectory evaluations. The bottleneck arises from the need to evaluate special functions such as elliptic integrals which can be slow to compute. 

To mitigate this cost, we employ the post-Newtonian (PN) expansion of Kerr geodesics. In particular, we use the 3PN Fourier-series eccentricity expansions for bound orbits derived by \citet{Sago2015}, which express the coordinate motion $r(\lambda)$, $\theta(\lambda)$, $\phi(\lambda)$, and $t(\lambda)$ as analytic functions of Mino time, without requiring computation of special functions. These expansions exploit the integrable structure of Kerr geodesics and are valid in the weak-field limit for moderate eccentricities and inclinations. This post-Newtonian framework offers orders-of-magnitude speedup over the exact method, making it well-suited for use in Bayesian inference algorithms. Each coordinate is given by a Fourier series expansion:

\begin{align}
    r(\lambda) &= p\sum_{n_r=0}^{\infty} \alpha_{n_r} \cos n_r\Upsilon_r\lambda, \\
    \cos\theta(\lambda) &= \sqrt{1-Y^2} \sum_{n_\theta=0}^\infty \beta_{n_\theta} \sin n_\theta \Upsilon_\theta \lambda, \\
    t(\lambda) &= \Upsilon_t\lambda + t^{(r)}(\lambda) + t^{(\theta)}(\lambda), \\
    \phi(\lambda) &= \Upsilon_\phi\lambda + \phi^{(r)}(\lambda) + \phi^{(\theta)}(\lambda)
\end{align}
where $Y=\cos{i}$ and the coefficients $t^{(A)}, \phi^{(A)}$ are given by:
\begin{align}
    t^{(A)}(\lambda) &\equiv \sum_{n_A=1}^\infty \tilde{t}_{n_A}^{(A)}\sin n_A\Upsilon_A \lambda, \\
    \phi^{(A)}(\lambda) &\equiv \sum_{n_A=1}^\infty \tilde{\phi}_{n_A}^{(A)}\sin n_A\Upsilon_A \lambda.
\end{align}
The full expressions for the Fourier coefficients $\alpha_{n_r}$, $\beta_{n_\theta}$, $\tilde{t}_{n_A}^{(\{r,\theta\})}$, $\phi^{(\{r,\theta\})}(\lambda)$, and Mino-time frequencies $\Upsilon_{\{t,r,\theta,\phi\}}$, in terms of the orbital parameters $\{p, e, \chi_\bullet, i\}$, are given in Appendices A and B of \cite{Sago2015} up to sixth order in $e$.

The increased computational efficiency comes at the cost of accuracy in certain regions of parameter space, particularly for highly eccentric orbits or those deep in the strong-field regime (i.e., small semi-latus rectum $p$). The PN formalism remains sufficiently accurate for orbits with $p \gtrsim 50$ and moderate eccentricity $\lesssim 0.5$, which encompass the parameter range relevant for most QPEs. We explore the accuracy of our 3PN computations, and the limits in which their validity breaks down, in Appendix~\ref{subsec:pn_vs_kerr}.

\subsection{Rigidly precessing disk} \label{sec:disk}\label{sec:likelihood}

\begin{nolinenumbers}
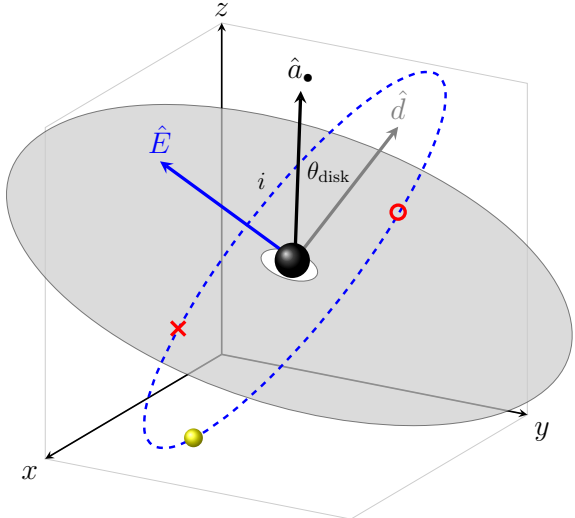
\begin{figure}
\centering
\resizebox{0.5\textwidth}{!}{
\tdplotsetmaincoords{70}{120}
\begin{tikzpicture}[tdplot_main_coords, scale=2]
\tikzset{every node/.style={font=\Large}}
\draw[black, opacity=0.2] (0,0,0) -- (3,0,0) -- (3,3,0) -- (0,3,0) -- cycle;
\draw[black, opacity=0.2] (0,0,0) -- (0,0,3) -- (0,3,3) -- (0,3,0) -- cycle;
\draw[black, opacity=0.2] (0,0,0) -- (3,0,0) -- (3,0,3) -- (0,0,3) -- cycle;

\draw[thick,-stealth] (0,0,0) -- (3,0,0) node[below left]{$x$};
\draw[thick,-stealth] (0,0,0) -- (0,3,0) node[below right]{$y$};
\draw[thick,-stealth] (0,0,0) -- (0,0,3) node[above]{$z$};

\coordinate (diskcenter) at (1.5, 1.53, 1.56);
\pgfmathsetmacro{\diskRotAlpha}{90}
\pgfmathsetmacro{\diskRotBeta}{acos(3/sqrt(10))}
\pgfmathsetmacro{\diskOuterRadius}{2.5}
\pgfmathsetmacro{\diskInnerRadius}{0.25}

\tdplotsetrotatedcoords{\diskRotAlpha}{\diskRotBeta}{0}

\begin{scope}[tdplot_rotated_coords, shift={(diskcenter)}]
    \fill[gray!40, opacity=0.7] (0,0,0) circle (\diskOuterRadius);
    \fill[white] (0,0,0) circle (\diskInnerRadius);
    \draw[black, thin, opacity=0.5] (0,0,0) circle (\diskOuterRadius);
    \draw[black, thin, opacity=0.5] (0,0,0) circle (\diskInnerRadius);
\end{scope}

\begin{scope}[tdplot_main_coords]
    \draw[very thick, dashed, blue] plot[smooth, domain=0:360] 
    ({1.7+1.17*1.7*cos(\x)}, {1.7+1.17*0.8*sin(\x)}, {1.7+1.17*1.5*sin(\x)});
\end{scope}

\begin{scope}[shift={(-0.24cm, -0.71cm)}, x={(1cm,0cm)}, y={(0cm,1cm)}]
    \shade[ball color=yellow] (0,0) circle (0.08);
\end{scope}

 \begin{scope}[shift={(1.5cm, 1.21cm)}, x={(1cm,0cm)}, y={(0cm,1cm)}]
  \draw[red, ultra thick] (0,0) circle (0.06);
\end{scope}

 \begin{scope}[shift={(-0.37cm, 0.22cm)}, x={(1cm,0cm)}, y={(0cm,1cm)}]
  \draw[red, ultra thick] (-0.06,-0.06) -- (0.06,0.06);
  \draw[red, ultra thick] (-0.06, 0.06) -- (0.06,-0.06);
\end{scope}

\begin{scope}[tdplot_main_coords]
    \coordinate (ellipse-center) at (1.7, 1.7, 1.7);
    
    \draw[ultra thick, -stealth, blue] (ellipse-center) -- ++(1.125, -0.675, 1.125) node[above] {\Large $\hat E$};
\end{scope}

\begin{scope}[tdplot_main_coords]
    \coordinate (ellipse-center) at (1.7, 1.7, 1.7);
    
    \draw[ultra thick, -stealth, black] (ellipse-center) -- ++(-0.1,0, 1.5) node[above] {\Large $\hat{a}_\bullet$};
\end{scope}

\begin{scope}[tdplot_main_coords]
    \coordinate (ellipse-center) at (1.7, 1.7, 1.7);
    
    \draw[ultra thick, -stealth, gray] (ellipse-center) -- ++(2.4,2.4, 2.4) node[above] {\Large $\hat{d}$};
\end{scope}

\begin{scope}[shift={(0.6cm, 0.8cm)}, x={(1cm,0cm)}, y={(0cm,1cm)}]
    \shade[ball color=black] (0,0) circle (0.15);
\end{scope}

\node[below left] at (0,0.5,1.8) {\large $i$};

\node[below left] at (0,1.33,2.05) {\large $ \theta_{\rm disk}$};

\end{tikzpicture}}
\caption{The blue dashed ellipse represents the orbit of the secondary body, inclined relative to the accretion disk (gray plane). The vectors $\hat{E}$, $\hat{a}_\bullet$, and $\hat{d}$ denote the orbit normal, black hole spin axis, and disk normal, respectively. The angles $i$ and $\theta_{\rm disk}$ describe the tilts of the orbital and disk planes relative to the black hole spin axis. Note that generic Kerr orbits do not lie in a single plane but rather occupy a toroidal region. The inclination angle $i$ depicted here is a conserved quantity derived from the constants of motion, rather than a geometric angle between a fixed orbital plane and the black hole spin axis. The figure is schematic and not meant to represent the full three-dimensional structure of the orbit. The red circle and red cross mark the two points where the orbit intersects the disk: the circle denotes the intersection from below, and the cross denotes the intersection from above. The origin of the coordinate system is at the MBH.}
\label{fig:schematic}
\end{figure}

\end{nolinenumbers}

\begin{figure}
\hspace{-3mm}
    \includegraphics[width=1.05\linewidth]{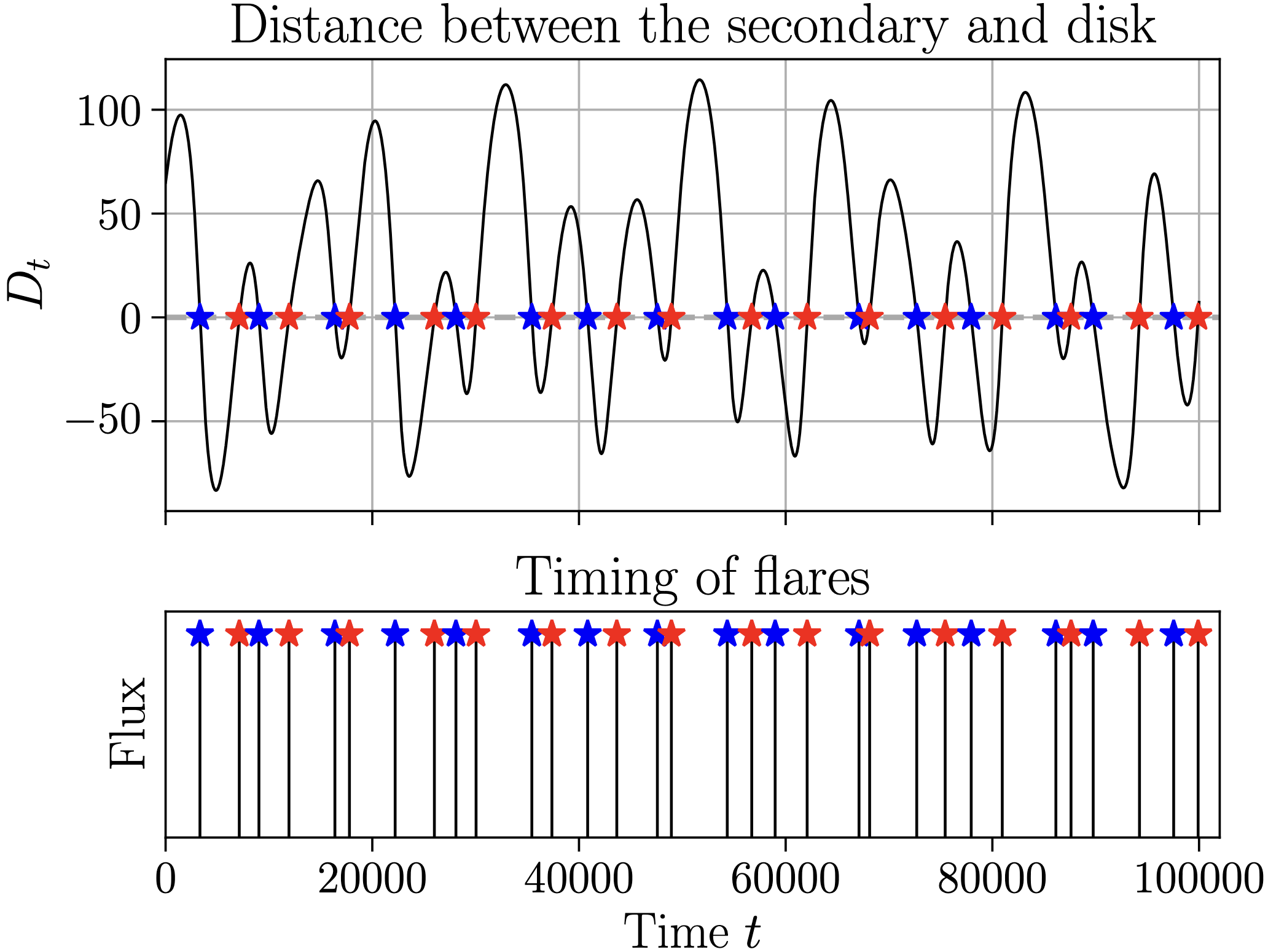}
    \caption{Evolution of the distance between the secondary object and the accretion disk, and the corresponding timing of flares. The top panel shows the distance between the secondary and the disk as a function of time, with flare events indicated by red and blue stars when the distance crosses zero. The bottom panel shows the extracted flare timings, where each vertical line marks the time of a disk crossing and the color distinguishes successive crossings coming from above and below the disk. The flares shown in the bottom panel do not have flux units; their heights are purely schematic. Only the timing of the flares conveys quantitative information in this plot. Distance and time are in geometric units with $R_g=GM_\bullet/c^2$ and $t_g=GM_\bullet/c^3$.  The semi-major axis is 100 $R_g$, eccentricity is 0.3, $i=40^\circ$, $\chi_\bullet=0.4$, $\theta_{\rm disk}=30^\circ$ and $T_{\rm disk}=1000 t_g$.}
    \label{fig:distanceplot}
\end{figure}

Equipped with the orbital trajectory of the secondary, the natural next step is to compute its crossing times for a static accretion disk, under the simplifying assumption that it is aligned with the MBH spin axis. In fact, this was explored in several early QPE timing studies \citep{Xian2021,Zhou2024a,Zhou2024b}. However, after the first few observational studies of QPE timings, it quickly became evident that the simple-case static disk model is an incomplete description in at least some QPEs. Some systems display significant, unexpected irregularity in their burst timings \citep{Arcodia2022,Giustini2024} not immediately reconciled with model predictions. Closer looks in \cite{Chakraborty2024} and \cite{Miniutti2025} observed super-periodic structure in the QPE arrival times of eRO-QPE1 and GSN 069, respectively, suggesting the presence of a further modulating timescale not present in the static disk model.

One promising avenue to explain these super-orbital modulations is rapid disk precession. As QPEs are sometimes associated with the aftermath of TDEs \citep{Chakraborty2021,Chakraborty2025a,Quintin2023,Miniutti2023a,Nicholl2024}, and TDE accretion disks are generally formed with arbitrary initial inclination, a significant disk inclination with respect to the MBH spin axis is not surprising. Moreover, the chaotic fluid dynamics of tidal stream circularization, stream-stream collisions, and accretion flow formation in Kerr spacetime further complicate the relationship between disk, stellar orbit, and MBH inclinations, further motivating the presence of misaligned disks. Such disks will be subject to rapid nodal precession via the frame dragging effect \citep{Stone2012,Franchini2016}, which in turn modulates the orbiter-disk collision timings. The precessing disk is bound to align on a timescale dependent on black hole mass and effective viscosity, which can be as long as several years \citep{Franchini2016}. Previous modeling efforts have already laid the groundwork for modeling the effect of an inclined, rigidly precessing accretion disk on QPE timings \citep{Franchini2023,Zhou2024c}; we build upon it here with our efficient Bayesian inference code.

We adopt a coordinate system with orthonormal basis $(\hat{x},\,\hat{y},\,\hat{z})$ where the spin axis of the MBH, $\hat{a}_\bullet$, is aligned with the $\hat{z}$-axis ($\hat{z} = \hat{a}_\bullet$). The origin of the coordinate system is also chosen to coincide with the MBH. In this configuration, the normal vector of the accretion disk, $\hat{d}$, is inclined at an angle $\theta_{\rm disk}$ relative to $\hat{a}_\bullet$. Similarly, the normal vector to the secondary's orbital plane, $\hat{E}$, is inclined relative to $\hat{a}_\bullet$ at an angle $i$. This coordinate system is particularly convenient because both the secondary's orbit and the accretion disk undergo nodal precession about the $\hat{z}$-axis. As a result, only two components of $\hat{d}$ evolve with time.

The disk normal vector as a function of time is given by:
\begin{align}
\hat{d}(t) = \begin{bmatrix}
    \sin\theta_{\rm disk} \cos\Big(2\pi t/T_{\rm disk} + \phi_{{\rm disk},0}\Big) \\
    \sin\theta_{\rm disk} \sin\Big(2\pi t/T_{\rm disk} + \phi_{{\rm disk},0}\Big) \\
    \cos\theta_{\rm disk}
\end{bmatrix},
\end{align}
where $T_{\rm disk}$ is the Lense-Thirring (nodal) precession period of the accretion disk, and $\phi_{{\rm disk},0}$ is an arbitrary initial phase.

The EMRI orbit is parameterized as $\vec{r}(t) = (x(t), y(t), z(t))$, which represents the full geodesic trajectory. This trajectory incorporates the complexities of apsidal precession, nodal precession, and orbital evolution. The orbit of the secondary in relation to the primary black hole and the accretion disk is shown in Fig.\ \ref{fig:schematic}. The distance $D$ between the disk and the EMRI as a function of time is given by the dot product $\hat{d} \cdot \vec{r}(t)$:
\begin{align}
\label{eq:D_t}
D(t) &= \sin\theta_{\rm disk} \cos\bigg(2\pi t/T_{\rm disk} + \phi_{{\rm disk},0}\bigg) x(t) \nonumber \\&+ \sin\theta_{\rm disk} \sin\bigg(2\pi t/T_{\rm disk} + \phi_{{\rm disk},0}\bigg) y(t)\nonumber \\&+ \cos\theta_{\rm disk} \, z(t).
\end{align}
The zeros of \(D(t)\) correspond to the QPE timings. $D(t)$ can take positive or negative values, indicating which side of the disk the EMRI is located on at a given time. 

We note that a limitation of our timing model is the assumption of a constant offset between the orbiter-disk collision and the QPE peak timing. It is certainly possible that variations in the accretion disk properties, mutual inclination between orbiter and disk, and/or instantaneous orbital velocity may result in differing rise timescales between eruptions. Moreover, recent work has suggested QPE may be powered by stellar debris-disk collisions \citep{Yao2025a,Linial2025} which would further complicate the timings. However, given the considerable uncertainty regarding which physical effects dominate and to what extent, the lack of a complete theoretical understanding of how flare profiles relate to collision dynamics, and the generally constant rise times observed in any given QPE (within a factor of a few), we assume a constant offset for simplicity.

\subsection{Likelihood for QPE timings} 
The likelihood function quantifies how well the predicted disk-crossing times match the observed QPE timings. These predicted crossings are determined by the zeros of the disk-crossing distance function $D(t, \vec{\xi})$, previously defined in Eq.~\ref{eq:D_t}, where $\vec{\xi}$ denotes the full set of model parameters. The disk precession parameters $(\theta_{\rm disk}, T_{\rm disk}, \phi_{{\rm disk},0})$ describe the time-dependent orientation of the accretion disk: $\theta_{\rm disk}$ is the inclination of the disk relative to the MBH spin axis, $T_{\rm disk}$ is the disk precession period, and $\phi_{{\rm disk},0}$ is the initial phase of the disk orientation.

The components of the secondary’s trajectory, $x(t)$, $y(t)$, and $z(t)$, are determined by bound Kerr geodesics and encapsulate relativistic effects such as apsidal and nodal precession. These trajectories depend on the orbital parameters $(\chi_\bullet, p, e, i, q_{r,0}, q_{z,0}, q_{\phi,0})$, which uniquely characterize the geodesic. Given a realization of model parameters $\vec{\xi}$, the function $D(t, \vec{\xi})$ can be evaluated over time, and its zeros correspond to the predicted flare times $t_{\text{model}, j}(\vec{\xi})$. For our code, we evaluate the zeros by identifying values of $t$ for which $D(t)$ changes sign, meaning the sampling resolution at which we evaluate $D(t)$ is important for inference accuracy. These computed disk crossings are then compared to the observed QPE timings $t_{\text{obs}, i}$ to compute the likelihood. Figure~\ref{fig:distanceplot} illustrates how the structure of $D(t, \vec{\xi})$ determines the timing of observable eruptions.

The observed QPE timings, compared to the orbital disk crossing times, are modified by the Shapiro and geometric (Roemer) delays. The Shapiro delay is given by:
\begin{align}
    \Delta_{\rm Shap} &= -2M_\bullet\ln[r(1+\hat{n}_{\rm obs}\cdot \hat{n}_{\rm cross})],
\end{align}
where $\hat{n}_{\rm cross}$ is the unit vector pointing from the black hole to the orbiter–disk collision location, as defined in Eq.~\ref{eq:D_t}, and $\hat{n}_{\rm obs}$ is the line-of-sight unit vector pointing from the black hole to the observer. We assume a fixed azimuthal viewing angle without loss of generality, such that the observer lies in the $x$–$z$ plane. The line of sight is then specified by a single polar angle $\theta_{\rm obs}$ measured from the $z$-axis:
\begin{align}
\hat{n}_{\rm obs} = \sin\theta_{\rm obs}\hat{x} + \cos\theta_{\rm obs}\hat{z}.
\end{align}
The geometric delay is then given by:
\begin{align}
    \Delta_{\rm geom} = -\vec{r}_{\rm cross}\cdot \hat{n}_{\rm obs}.
\end{align}
Therefore, the complete set of parameters in our model is: 
\begin{align}
\vec{\xi} = \{&p, e, i, \chi_\bullet,M_\bullet,\theta_{obs},\theta_{disk},T_{disk};\nonumber \\&  q_{r,0},q_{z,0},q_{\phi,0},\phi_{\rm{disk},0}\},
\end{align}
where the parameters $(q_{r,0},q_{z,0},q_{\phi,0},\phi_{\rm{disk},0})$ are choices of initial phase which do not carry interesting physical information. For the purposes of the simulated injections in this study, we fix these initial phases to zero. It is possible that the initial phases act as nuisance parameters affecting the measurability of the other physically interesting parameters, particularly for observing times short compared to the precession timescales, but our analysis is not sensitive to this effect. In the next section, we convert the semilatus rectum $p$ to the semi-major axis\footnote{The semi-major axis $a$ should not to be confused with $\chi_\bullet$, which is the spin of the massive black hole. The semi-major axis $a$ is a commonly used orbital parameter in QPE literature, whereas the EMRI gravitational wave community typically describes the secondary's orbit using the semilatus rectum $p$.} $a=p/(1-e^2)$.

For a given window of continuous observations, the $i$th observed QPE is then matched against the timing of the $i$th computed disk crossing:
\begin{align}
r_i(\vec{\xi},t_{{\rm obs},i}) =  \left| t_{\text{obs}, i} - t_{\text{model}, i}(\vec{\xi}) \right|,
\label{eq:23}
\end{align}
where \(t_{\text{model}, i}(\vec{\xi})\) are the zeros of \(D(t, \vec{\xi})\), modified by the Shapiro and geometric delays, for a given set of parameters. To account for observational uncertainties, we also introduce a vector of timing errors $\{\sigma_i\}$ corresponding to the measurement uncertainty on each of the QPE timings. For our simulated data, we assume a constant error of $\sigma_i=100\;\rm sec$ for all QPEs (though with constant errors, any choice would be equivalent, as it would uniformly scale the likelihoods). We then divide the squared residuals of Eq.~\ref{eq:23} by the squared error to compute the \(\chi^2\) statistic across each of $N_{\rm obs}$ total observing windows:
\begin{align}
\chi^2(\vec{\xi}) = \sum_{j}^{N_{\text{obs}}} \sum_{i=1}^{n_{\text{qpe}},j} \left( \frac{r_i(\vec{\xi})}{\sigma_i} \right)^2.
\end{align}
The likelihood function is then expressed as:
\begin{align}
\mathcal{L}_z(\vec{\xi}) \propto \exp\left(-\frac{1}{2} \chi^2(\vec{\xi})\right),
\label{eq:25}
\end{align}
In this framework, the predicted disk-crossing times \(t_{\text{model}, i}(\vec{\xi})\) explicitly depend on the model parameters, via the trajectory of the orbiter. By minimizing the residuals between the observed and predicted timings, the likelihood assigns higher probabilities to parameter combinations that better match the data. In performing inference with this likelihood function, we assume flat priors such that the posterior is proportional to the likelihood via Bayes' theorem. We note that while geometric considerations suggest uniform priors in $\cos\theta$ for the angular parameters $\theta_{\rm obs}$ and $\theta_{\rm disk}$, we adopt flat priors in $\theta$ for simplicity. This choice has minimal impact for $\theta_{\rm obs}$ given its weak effect on timings (Appendix~\ref{sec:obsangle}), though the appropriate prior for $\theta_{\rm disk}$ remains uncertain and depends on the unknown formation/evolution mechanisms of QPE-hosting disks.

We performed Bayesian inference using the affine-invariant Markov Chain Monte Carlo (MCMC) ensemble sampler \citep{Goodman2010} as implemented in the \texttt{emcee} Python package \citep{DFM2013}. These calculations were carried out within our inference framework, \qpefit, which features trajectory and likelihood computations that are parallelized and GPU-accelerated using the \texttt{CuPy} Python library \citep{cupy}. \texttt{CuPy} provides a Python interface to CUDA, enabling high-throughput numerical computations on GPUs. Computations were performed on NVIDIA A100 GPUs available on the \texttt{engaging} cluster operated by the MIT Office of Research Computing and Data. We present the main results of our MCMC inference in Section~\ref{sec:results}. In addition to those results, we performed a large grid of simulations to explore the parameter space extensively, presented in Appendix~\ref{appendix:all_sim}.

\section{Results} \label{sec:results}
Using \qpefit\ (Section~\ref{sec:model}), we explored a large grid of synthetic QPE timings across the parameter space of $\{a, e, i, M_\bullet, \chi_\bullet, \theta_{\rm disk}, T_{\rm disk}\}$, also varying the observing strategy, to explore the ability of MCMC sampling to recover the true parameters. For each sampling run we inject a solution computed with the \textit{exact} Kerr geodesic trajectory \citep{kerrgeopy}, then perform inference with the less accurate, but more efficient, PN sampler. We begin with only the static disk model; we then generalize to a misaligned precessing disk to show its confounding effect on parameter inference; and finally apply our model to the particularly irregular QPE timing data of eRO-QPE1 to showcase its ability to describe observations of even the erratic sources, albeit with extremely multimodal posteriors.

\subsection{The effects of varying semimajor axis, eccentricity, spin, and $M_\bullet$ with a non-precessing disk} \label{subsec:no_precess}

\begin{figure*}
    \centering
    \includegraphics[width=\linewidth]{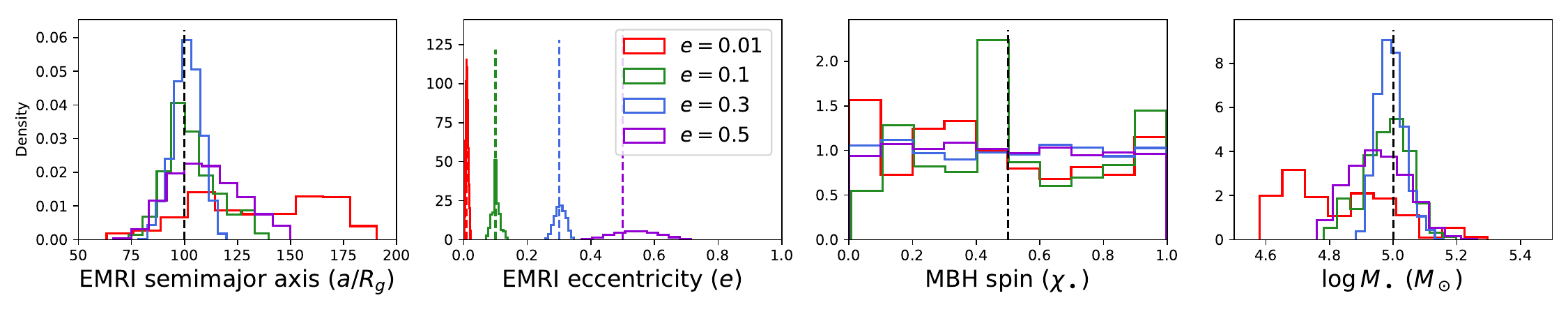}
    \includegraphics[width=\linewidth]{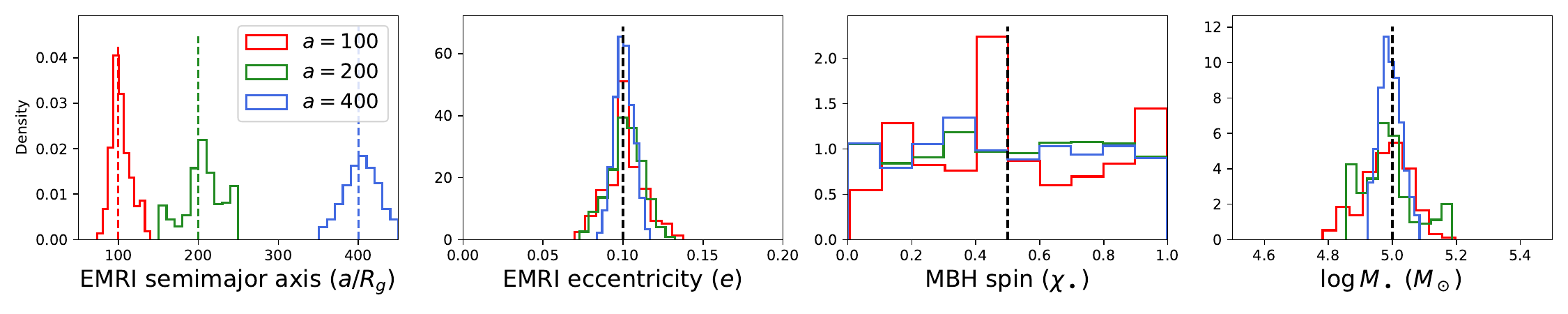}
    \includegraphics[width=\linewidth]{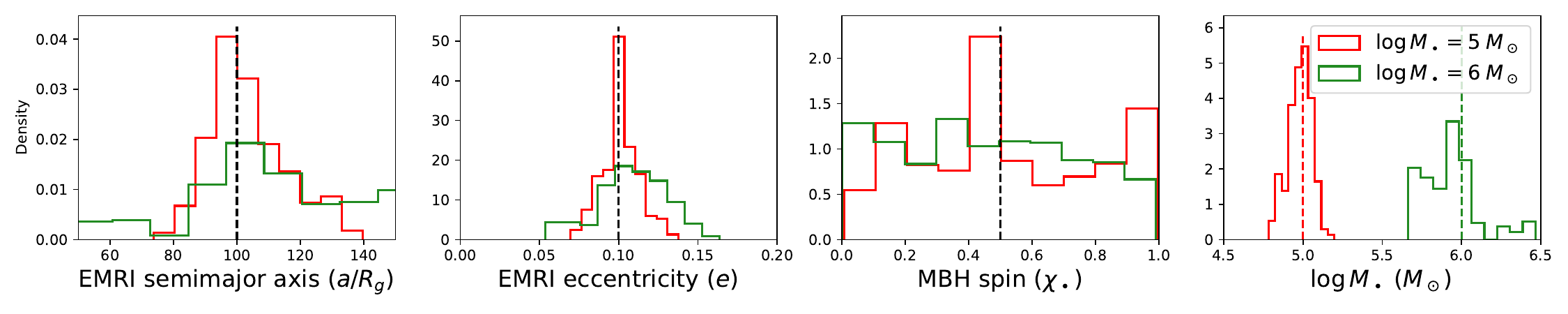}
    \caption{Parameter recovery in the static accretion disk case. \textbf{Top panel:} varying $e$ drawn from $e \in [0.01, 0.1, 0.3, 0.5]$, for a fixed semimajor axis of $a=100 R_g$, $M_\bullet=10^5 M_\odot$, and MBH spin $\chi_\bullet=0.5$. \textbf{Middle panel:} varying semimajor axes drawn from $a \in [100,200,400]\;R_g$, for a fixed eccentricity of $e=0.1$, $M_\bullet=10^5 M_\odot$, and MBH spin $\chi_\bullet=0.5$. \textbf{Bottom panel:} varying MBH mass drawn from $\log M_\bullet \in [5,6]\;M_\odot$, for a fixed semimajor axis of $a=100\;R_g$, eccentricity of $e=0.1$, and MBH spin $\chi_\bullet=0.5$.}
    \label{fig:vary_noprecess}
\end{figure*}

We first consider the effect of varying EMRI/MBH parameters in the simplest-case model, i.e. a static planar accretion disk aligned with the MBH spin axis. All results from the MCMC sampling runs reported in Sections~\ref{subsec:no_precess}-\ref{subsec:with_precess} were carried out with 1000 independent walkers for 100,000 steps, and used an uninterrupted 100~ks observing window. We use a large walkers-to-steps ratio because of the multimodality of the parameter space, with starting positions chosen from a uniform random distribution outlined in Table~\ref{tab:priors}. We discarded the first 50\% of steps as burn-in, then retained the 100,000 highest-likelihood points from the resulting distributions. We discuss the convergence of these MCMC chains in Appendix~\ref{appendix:convergence}. Given the broad, conservative priors, compatible with actual uncertainties on the underlying system properties given real data, it is remarkable that several of the parameters (e.g. $a$, $e$, $M_\bullet$) converge to the true values with posterior uncertainties of $\sim 10\%$.

\begin{table}
\caption{Prior ranges used for our Bayesian inference scheme.} 
\label{tab:priors}
\centerline{
\begin{tabular}{lcc}
\toprule
Parameter & Model \\
\midrule
Semi-major axis ($a$) & $\mathcal{U}(30,500)R_g$ \\
Eccentricity\footnote[2]{Note that while our prior range for eccentricity extends to $e = 0.9$, we only inject lower eccentricity signals using the exact Kerr geodesic model. The broader prior ensures we can assess potential biases when recovering these injections with a PN-approximated model.} ($e$) & $\mathcal{U}(0,0.9)$ \\
Inclination ($i$) & $\mathcal{U}(0^\circ,90^\circ)$ \\
MBH Mass [$\log(M_\bullet/M_\odot)$] & $\mathcal{U}(4,8)$ \\
Observer viewing angle ($\theta_{\rm{obs}}$) & $\mathcal{U}(0,\pi)$ \\
Disk inclination $(\theta_{\rm disk})$ & $\mathcal{U}(0^\circ,60^\circ)$ \\
Disk precession period ($T_{\rm disk}$) & $\mathcal{U}(2,100)P_{\rm orb}$ \\
\bottomrule
\end{tabular}}
\end{table}

In our first example, we inject a signal with semimajor axis $a=100R_g$ (where $R_g\equiv GM_\bullet/c^2$), detector-frame MBH mass $M_\bullet = 10^5 M_\odot$, and MBH spin $\chi_\bullet = 0.5$. We perform inference for four different choices of eccentricity, $e\in\{0.01, 0.1, 0.3, 0.5\}$. Our results are shown in the top panel of Fig.~\ref{fig:vary_noprecess}.  We observe that very low-eccentricity EMRIs ($e\sim0.01$) will not be as powerful for $M_{\bullet}$ mass constraints, as the timing modulation effects of apsidal precession are not as pronounced as in EMRIs with modest eccentricity ($e\sim0.1-0.3$). On the other hand, at higher eccentricities ($e\sim0.5$) the parameter estimation once again becomes worse, with wider errors on $a$, $e$, and $\log M_\bullet$. However, this is likely due to the breakdown of the PN eccentricity expansion used in our orbital trajectory calculation, rather than an intrinsically larger degeneracy. The MBH spin ($\chi_\bullet$) is unconstrained in all cases we explored, which are limited to relatively wide orbits ($\geq 100R_g$) and short observing baselines. 

\begin{figure*}
    \centering
    \includegraphics[width=0.65\linewidth]{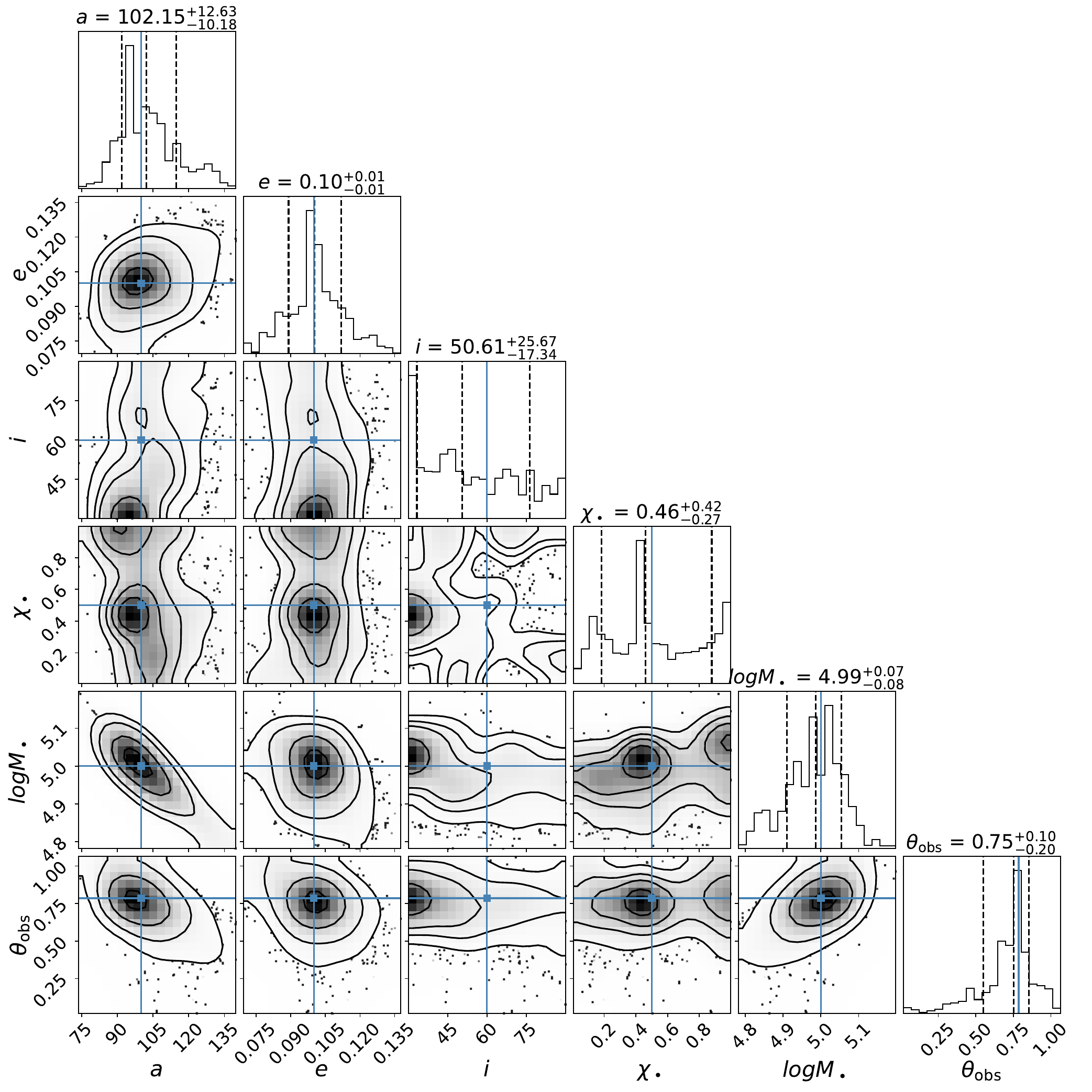}
    \caption{Corner plot for a static accretion disk aligned with the MBH spin axis. The posterior distributions are produced using \texttt{QPE-FIT} and the blue lines indicate the true injected values used to generate the synthetic QPE timing data. 2D posteriors are smoothed by a $2\sigma$ Gaussian kernel for visual clarity.}
    \label{fig:corner_noprecess}
\end{figure*}

For the middle panel of Fig.~\ref{fig:vary_noprecess}, we inject a signal with $e=0.1$, $M_\bullet=10^5 M_\odot$, $\chi_\bullet=0.5$, and three different choices of $a\in\{100,200,400\}R_g$. Remarkably, varying $a$ has relatively little effect on measurements of $M_\bullet$. While larger $a$ results in a slower apsidal precession (thus making $M_\bullet$ harder to measure), it also results in larger-amplitude timing modulations; we posit that the two effects offset each other to result in a similar overall mass constraint.

For the bottom panel of Fig.~\ref{fig:vary_noprecess}, we inject a signal with $a=100R_g$, $e=0.1$, $\chi_\bullet=0.5$, and two different choices of $\log (M_\bullet/M_\odot)\in\{5,6\}$. The effect of $M_\bullet$ enters predominantly through the QPE recurrence timescale: for a fixed $a/R_g$, the orbital timescale increases linearly with $M_\bullet$. Increasing $M_\bullet$ by $100\times$ thus requires a (usually infeasible) corresponding $100\times$ longer observing baseline, making lower-mass MBHs better-suited for QPE discovery and parameter estimation.

In Fig.~\ref{fig:corner_noprecess}, we show a corner plot from a sampling run with injected parameters of $a=100R_g$, $e=0.1$, $M_\bullet=10^5M_\odot$, $\chi_\bullet=0.5$, $i=60^\circ$, and $\theta_{\rm obs}=\pi/4$. The orbital semimajor axis is significantly degenerate with $M_\bullet$ (as expected, see e.g. Fig. 4 of \citealt{Kejriwal2025}), and the main effect disentangling them is the timescale of Schwarzschild (apsidal) precession. Eccentricity is essentially well-constrained within just a few bursts, whereas the EMRI orbital inclination ($i$) and MBH spin ($\chi_\bullet$) are completely unconstrained. The former is due to the relatively large radii at which QPEs occur ($a \gtrsim 100R_g$): the Kerr metric is not spherically symmetric, so spacetime does vary with $i$, but only weakly except for strong-field orbits.

The dominant observable effect of the black hole spin $\chi_\bullet$ is Lense-Thirring (nodal) precession, but only over timescales significantly longer than the ones probed by our experiments. Long-term dissipative effects are also likely to set in over these baselines, but our analysis does not incorporate these; in follow-up work, we will examine parameter estimation over much longer baselines while accounting for secular evolution due to interaction with the environment. Other QPE timing studies, e.g. \cite{Xian2021} and \cite{Zhou2025}, have explored spin constraints by considering these longer baselines, and in some cases produce good constraints with uncertainties $<0.1$.

We also varied several of the other parameters, such as the observing angle $\theta_{\rm obs}$ and MBH spin ($\chi_\bullet$), and found that they had limited impact on the parameter recovery. See Appendix \ref{sec:obsangle} for further discussion.

\subsection{The effects of an inclined, rigidly-precessing disk} \label{subsec:with_precess}
\begin{figure}
    \centering
    \includegraphics[width=0.9\linewidth]{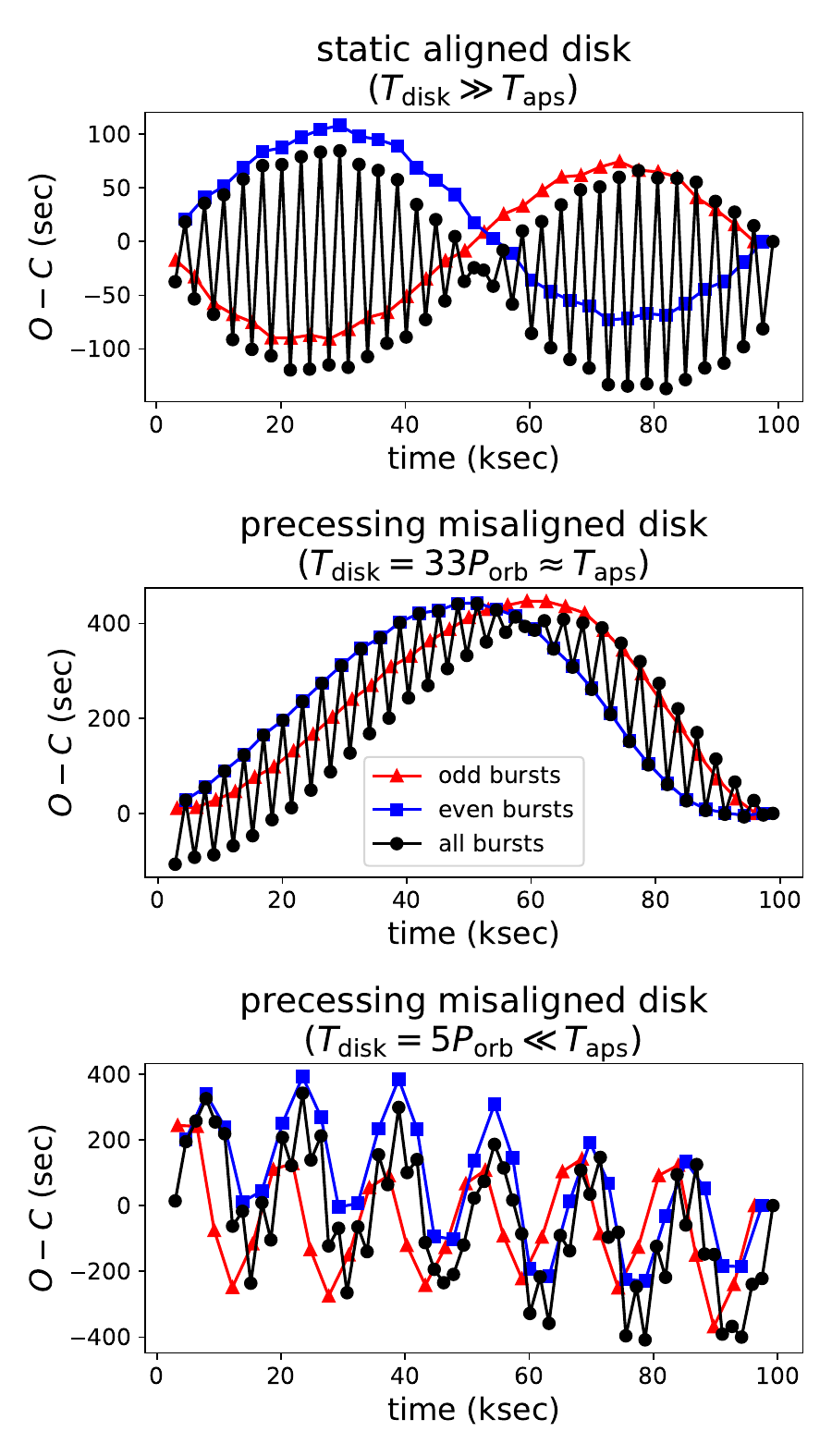}
    \caption{Synthetic $O-C$ diagrams for an EMRI with $a=100R_g$, $e=0.1$ around an MBH of $M_\bullet=10^5M_\odot$, $\chi_\bullet=0.5$. In the \textbf{top panel}, we show the simple case of a static disk with no precession or misalignment. We then introduce slow (\textbf{middle panel}) and rapid (\textbf{bottom panel)} disk precession, with an inclination of $\theta_{\rm disk}=20^\circ$. We construct $O-C$ diagrams considering all bursts together (black circles), odd bursts only (red triangles), and even bursts only (blue squares). }
    \label{fig:oc}
\end{figure}

As discussed in Section~\ref{sec:disk} and in prior studies \citep{Franchini2023,Linial2024b,Zhou2024c}, the static in-plane disk model of Section~\ref{subsec:no_precess} appears inconsistent with observations in some QPEs. Systems such as eRO-QPE1 \citep{Chakraborty2024} and GSN 069 \citep{Miniutti2025} show long-term modulations in QPE arrival times not predicted by static disk models, pointing instead to a further modulating timescale. One possible (but not unique) explanation is rapid disk precession, naturally expected for an isotropic distribution of TDEs with respect to the black hole spin axis. Given that the weak-field per-orbit apisdal motion is approximately $\delta\varphi \approx 6\pi (a/R_g)^{-1}(1-e^2)^{-1}$, the EMRI apsidal precession timescale ($\Delta\varphi=2\pi$) is given by:
\begin{equation}
    T_{\rm aps} \approx \frac{1}{3}\bigg(\frac{a}{R_g}\bigg)(1-e^2)P_{\rm orb}.
    \label{eq:T_aps}
\end{equation}
This expression occupies values between $\sim 25-133P_{\rm orb}$ for the range of eccentricities and semimajor axes explored by our trials. The predominant effect qualitatively impacting QPE timing behavior is how $T_{\rm aps}$ compares with $T_{\rm disk}$.

Before proceeding to injection/recovery trials, we first provide some illustrative examples to showcase these effects, for cases where $T_{\rm disk}\gg T_{\rm aps}$, $T_{\rm disk}\approx T_{\rm aps}$, and $T_{\rm disk}\ll T_{\rm aps}$. In Fig.~\ref{fig:oc} we show synthetic $O-C$ (observed minus calculated) diagrams, where the ``calculated'' timings assume a constant recurrence time and the ``observed'' timings are modified by the combined orbital and disk precessions. We construct $O-C$ diagrams considering all bursts together, as well as odd and even bursts separately, as this effectively isolates the orbital evolution of the EMRI subject to precession effects.

The two parameters affecting the deviation from the static, aligned-disk case are the disk inclination angle with respect to the MBH spin axis ($\theta_{\rm disk}$) and the disk precession period ($T_{\rm disk}$). In the top panel, we show the $O-C$ diagram expected from the static disk model (which is equivalent to taking $T_{\rm disk}\rightarrow \infty$), in which the odd and even QPE flares trace sinusoids in antiphase over the apsidal precession period (red/blue points) while also producing the characteristic ``long-short'' recurrence time alternation on shorter timescales (black points). For a slowly precessing disk, with $T_{\rm disk}=33P_{\rm orb}\approx T_{\rm aps}$, the apsidal precession is still measurable, and the characteristic long-short alternation is preserved (black points), but the odd and even bursts now vary together in phase (red/blue points). For sufficiently rapid disk precession, the long-short alternation is eliminated completely, with the disk precession becoming the dominant timing effect---in the bottom panel we show the case of $T_{\rm disk}=5P_{\rm orb}\ll T_{\rm aps}$, in which a sinusoidal trend on the disk precession timescale becomes the dominant effect. The middle panel is qualitatively similar to the $O-C$ diagram computed for GSN 069 in \cite{Miniutti2025}, in the sense that there is a long/short alternation on timescales $\approx P_{\rm orb}$ and an in-phase modulation on timescales $\gg P_{\rm orb}$, whereas the bottom panel is more akin to eRO-QPE1 \citep{Chakraborty2024}, suggesting a range of disk dynamical properties across the QPE population. No known QPEs have thus far been shown to be compatible with the static-disk picture.

\begin{figure*}
    \centering
    \includegraphics[width=\linewidth]{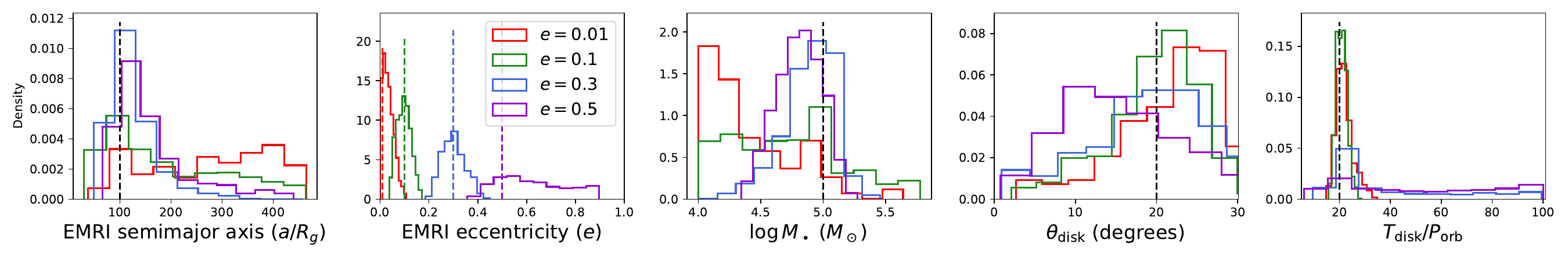}
    \includegraphics[width=\linewidth]{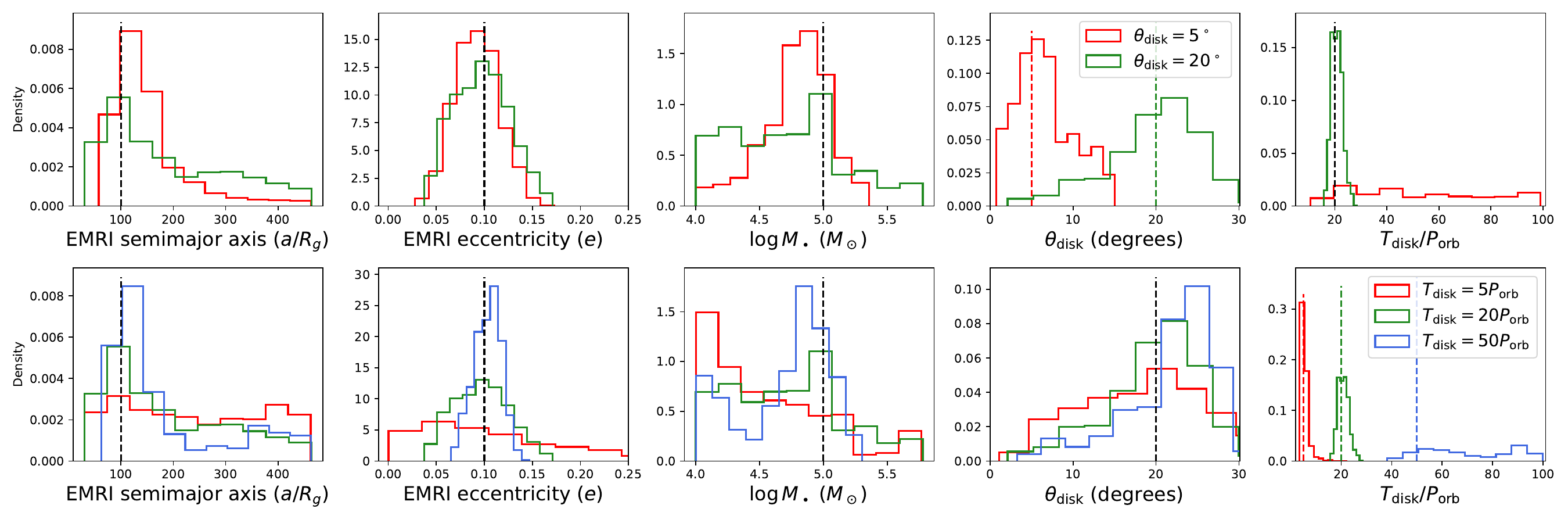}
    \caption{\textbf{Top panel:} Varying eccentricities drawn from $e \in [0.01, 0.1, 0.3, 0.5]$, for a fixed semimajor axis of $a=100 R_g$, $M_\bullet=10^5 M_\odot$, MBH spin $\chi_\bullet=0.5$, disk inclination $\theta_{\rm disk}=20^\circ$, and disk precession period $T_{\rm disk}=20P_{\rm orb}$. \textbf{Middle panel:} varying accretion disk inclination angle, $\theta_{\rm disk}\sim[5^\circ,20^\circ]$. \textbf{Bottom panel:} varying disk precession period $T_{\rm disk}\sim[5,20,50]P_{\rm orb}$.}
    \label{fig:vary_withprecess}
\end{figure*}

Following this toy example, we now proceed to injection/recovery runs including a misaligned rigidly precessing accretion disk. In the top panel of Fig.~\ref{fig:vary_withprecess} we show, by analogy to Fig.~\ref{fig:vary_noprecess}, the eccentricity constraints with a precessing disk for a 100~ks observing window (about 50 consecutive QPEs). We do not duplicate the posterior distribution of $\chi_\bullet$ because it is equally unconstrained compared to the non-precessing case. In the top panel of Fig.~\ref{fig:vary_withprecess} we inject solutions with $a=100R_g$, $M_\bullet=10^5M_\odot$, $\chi_\bullet=0.5$, $\theta_{\rm disk}=20^\circ$, $T_{\rm disk}=20P_{\rm orb}$, and four different eccentricities $e\sim\{0.01,0.1,0.3,0.5\}$. In the middle panel, we hold fixed $e=0.1$, use two different disk inclinations $\theta_{\rm disk}\in\{5^\circ,20^\circ\}$, and leave all other parameters the same. In the bottom panel, we fix $\theta_{\rm disk}=20^\circ$, use three different disk precession periods $T_{\rm disk}\in\{5,20,50\}P_{\rm orb}$, and leave other parameters unchanged. In the middle/bottom panels we vary the disk inclination angle $\theta_{\rm disk}$ and precession period $T_{\rm disk}$, respectively.

Relative to the static-disk scenario, the parameter constraints degrade significantly. This is an expected consequence of introducing additional degrees of freedom without augmenting the data with new constraining features such as light curve morphology or spectra. The observed sequence of QPE arrival times cannot in general break the degeneracies between multiple parameter combinations that produce similar timing modulations, at least for the baselines we examined. This inherent limitation underscores a key challenge in using QPE timing data alone to constrain EMRI models with realistic disk physics. Using additional observables or longer baselines may help to lift these degeneracies and better constrain the underlying physical parameters.

\begin{figure*}
    \centering
    \includegraphics[width=0.9\linewidth]{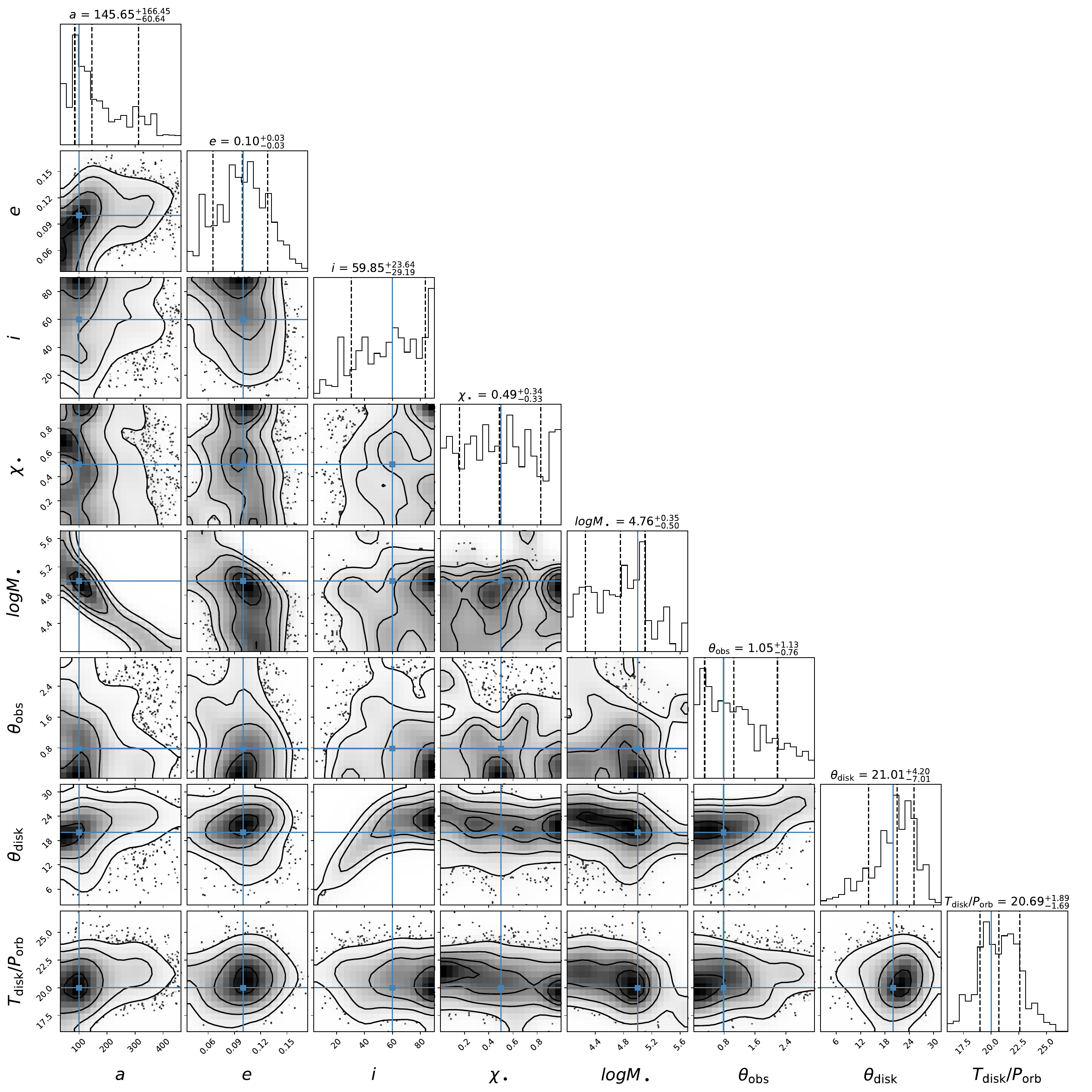}
    \caption{Corner plot for an inclined, rigidly precessing accretion disk. As in Fig.\ \ref{fig:corner_noprecess}, the posterior distributions are produced using \texttt{QPE-FIT} and the blue lines indicate the true injected values used to generate the synthetic QPE timing data. $a$ and $M_\bullet$ are more poorly constrained than in the static-disk case (Fig.~\ref{fig:corner_noprecess}), while $e$ is still recovered well, as are the disk parameters ($\theta_{\rm disk}$, $T_{\rm disk}$). 2D posteriors are smoothed by a $2\sigma$ Gaussian kernel for visual clarity.}
    \label{fig:corner_withprecess}
\end{figure*}

In Fig.~\ref{fig:corner_withprecess} we show a corner plot from a sampling run with $a=100R_g$, $e=0.1$, $M_\bullet=10^5M_\odot$, $\chi_\bullet=0.5$, $\theta_{\rm obs}=\pi/4$, $\theta_{\rm disk}=20^\circ$, and $T_{\rm disk}=20P_{\rm orb}$. There is a significant degeneracy between $i$ and $\theta_{\rm disk}$, and a slight correlation between $e$ and $\theta_{\rm disk}$. The increased parameter dimensionality significantly worsens constraints on the inference of most parameters, with only $e$, $\theta_{\rm disk}$, and $T_{\rm disk}$ appearing to converge. Given that disk precession (or some other modulating timescale) appears to be common, we advise longer observational baselines for parameter inference.

\subsection{The effect of neglecting disk precession when performing parameter inference} \label{subsec:bias}

\begin{figure*}
    \centering
    \includegraphics[width=\linewidth]{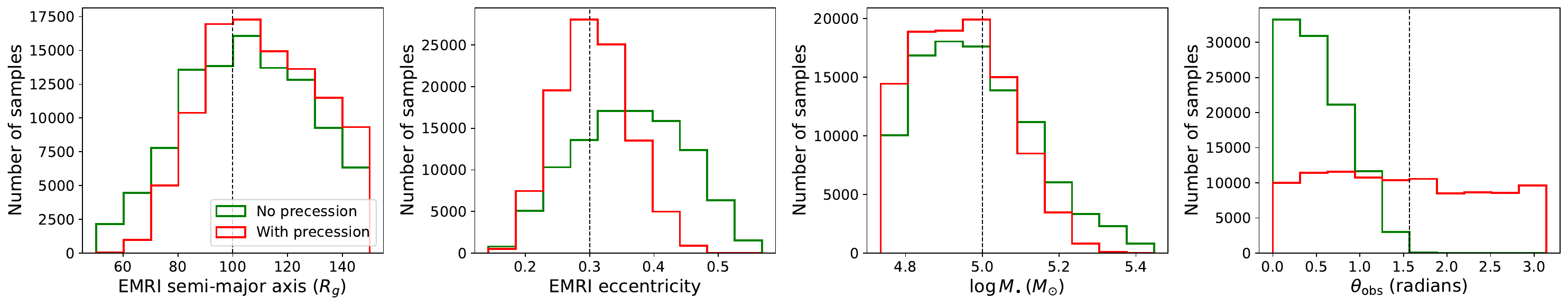}
    \caption{Parameter recovery when using a precessing versus non-precessing likelihood for an EMRI with $a=100R_g$ and $e=0.3$ around an MBH with $M_\bullet=10^5M_\odot$, $\chi_\bullet=0.5$. We inject a sequence of timings, assuming a disk with inclination $\theta_{\rm disk}=10^\circ$ and disk precession period $T_{\rm disk}=20P_{\rm orb}$ and an observing angle of $\theta_{\rm obs}=\pi/2$. We then recover parameters using a likelihood that assumes a precessing disk (red) or non-precessing disk (green).}
    \label{fig:precessionbias}
\end{figure*}

We further consider the effect of fitting observations assuming a static disk, when the true underlying process is more complex. Omitting precession leads to systematic biases across all parameters, driven by the model’s failure to capture key time-dependent structure. To evaluate the impact of this effect on parameter estimation, we inject a solution with a precessing disk, then perform inference with 500 walkers$\times$5000 steps \textit{assuming the static-disk model.} Figure~\ref{fig:precessionbias} shows the resulting posterior distributions, while Table~\ref{tab:bias} summarizes maximum a posteriori (MAP) values, uncertainties, and deviations from the true parameters

The largest bias appears in the observing angle, with a MAP shift to $\theta_{\rm obs} = 0.38$, nearly $4\sigma$ from the truth, reduced to $0.92\sigma$ when precession is included. Biases in $\log M_\bullet$, semi-major axis, and eccentricity are similarly reduced, with deviations falling below $0.2\sigma$ under the precessing model. Posterior widths also decrease, except for $\theta_{\rm obs}$, which remains weakly constrained due to its weak effect on the observed burst timings. These results underscore the importance of modeling disk precession to accurately recover system parameters in the presence of longer-period modulations in the signal. While acceptable fits can be sometimes be obtained with static-disk models over short baselines, longer baselines in particular require precession to be correctly modeled in order to avoid biased inferences. 

\begin{table*}
\caption{Recovered parameter statistics when using a precessing versus non-precessing likelihood for an EMRI with same parameters as Fig.\ \ref{fig:precessionbias}.}
\label{tab:bias}
\centerline{
\begin{tabular}{lccccc}
\toprule
Parameter & Model & Maximum a posteriori & Standard deviation & Deviation from truth \\
\midrule
\midrule
Semi-major axis $a$ & No Precession & 88.5 & 22.5 & -0.51 $\sigma$ \\
 & Precession    & 103 & 20.1  &\\
\midrule
Eccentricity $e$ & No Precession & 0.343 & 0.084 & 0.51 $\sigma$ \\
 & Precession    & 0.315 & 0.056 &  \\
\midrule
BH Mass $\log M_\bullet$ & No Precession & 5.08 & 0.15 & 0.54 $\sigma$ \\
& Precession    & 4.98 & 0.12 &  \\
\midrule
Observing angle $\theta_{\rm{obs}}$ & No Precession & 0.169 & 0.352 & -3.99 $\sigma$ \\
 & Precession    &  0.740 & 0.900 & \\
\bottomrule
\end{tabular}}
\end{table*}

\subsection{The effects of varying observational cadence and baseline}
We now consider how varying the observation strategy modifies parameter constraints. For a given set of parameters, observing more QPEs with higher density and longer baseline generally improves constraints, at the cost of more resources. To quantify the information gain of additional observing time, we performed sampling runs for three different observing strategies: $1\times 100$ kilosecond observation (chosen to approximately match the capability of \textit{XMM-Newton}); $4\times 100$ kilosecond observations, with short gaps between snapshots of 200 kiloseconds (roughly mimicking the observing strategy of four contiguous \textit{XMM-Newton} orbits); and $4\times 100$ kilosecond observations with long gaps of 1 megasecond between observations.

\begin{figure*}
    \centering
    \includegraphics[width=\linewidth]{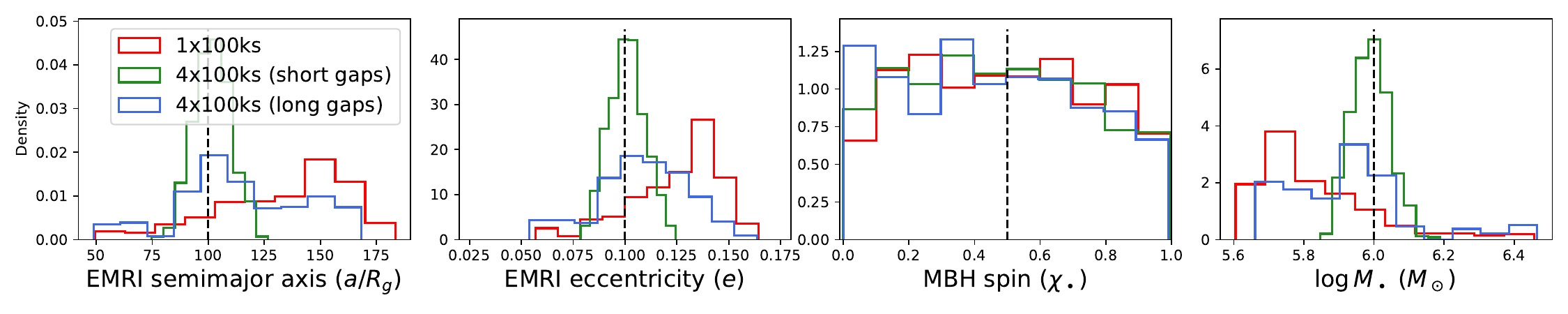}
    \includegraphics[width=\linewidth]{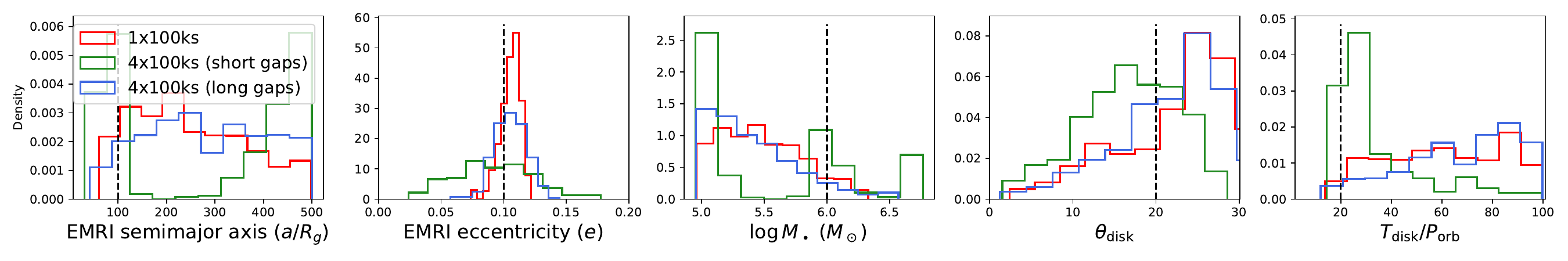}
    \caption{Parameter recovery for various observing strategies ($1\times 100$~ks, $4\times 100$~ks with 200~ks gaps, $4\times 100$~ks with 1~Ms gaps) for an EMRI with $a=100R_g$ and $e=0.1$ around an MBH with $M_\bullet=10^6M_\odot$, $\chi_\bullet=0.5$. 
    \textbf{Top panel:} no disk precession. \textbf{Bottom panel:} with disk precession, for an assumed $\theta_{\rm disk}=20^\circ$ and $T_{\rm disk}=20P_{\rm orb}$.} 
    \label{fig:vary_obs}
\end{figure*}

In Fig.~\ref{fig:vary_obs}, we show the effect of varying the invested observation time, first for an in-plane static disk (top panel). For an assumed system with $a=100R_g$, $e=0.1$, $M_\bullet=10^6M_\odot$ (note the higher-than-typical value chosen here), $\chi_\bullet=0.5$, we see that $1\times 100$ kilosecond observation---which observes 5 consecutive eruptions---provides weak constraints on the EMRI semimajor axis, eccentricity, and $M_\bullet$, due to the small number of bursts. Increasing to $4\times100$~ks observations results in improved constraints, but for the same invested time, we find that a \textit{shorter} baseline with \textit{smaller} observing gaps is better able to constrain $a$, $e$, and $M_\bullet$. This is a surprising result, as this strategy covers a smaller fraction of the overall apsidal period. We suggest this may be due to the difficulty of uniquely attributing a burst number following a longer gap. This is even more pronounced for the case including disk precession (bottom panel of Fig.~\ref{fig:vary_obs}), which further adds the ambiguity of uniquely modeling the apsidal motion.

\subsection{Application to data of eRO-QPE1} \label{subsec:application}

\begin{figure*}
    \centering
    \includegraphics[width=\linewidth]{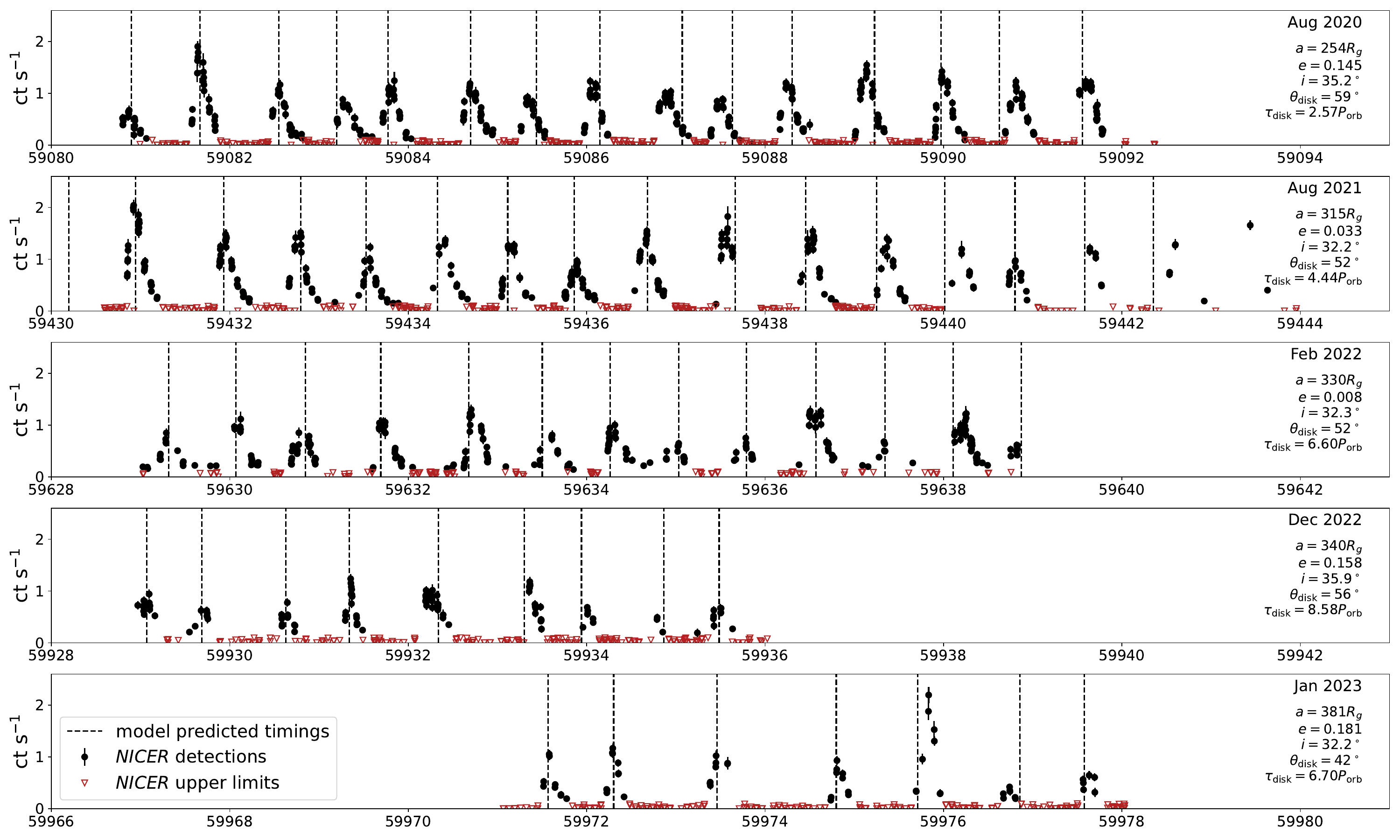}
    \caption{Our QPE timing model fit to the eruptions of eRO-QPE1. While the posteriors are extremely multi-modal, making parameter inference difficult and unreliable, we show the highest-likelihood parameter combinations here, showing that as a proof-of-concept, a rigidly precessing accretion disk appears able to explain the irregular burst arrival times observed in eRO-QPE1, despite the presence of significant degeneracies and multi-modal posteriors.}
    \label{fig:ero1}
\end{figure*}

To showcase the ability of our model to describe QPE timing data over short baselines, particularly for the more erratic sources which are difficult/impossible to analyze within the standard static-disk model, we fit the burst timings of eRO-QPE1, a particularly irregular source in both its timing properties and eruption profiles \cite{Arcodia2022}. Using the \textit{NICER} data from \cite{Chakraborty2024}, we fit each epoch separately due to the long (month- to year-long) gaps between observations and the non-inclusion of secular evolution effects in our timing code. We used uniform priors (Table~\ref{tab:priors}) with $a\sim[100, 500]R_g$, $e\sim[0, 0.5]$, $i\sim[0^\circ,90^\circ]$, $\theta_{\rm disk}\sim[0, 60^\circ]$, and $T_{\rm disk}\sim[2,100]P_{\rm orb}$. We also added two additional parameters: $t_0\sim[-50000,50000]\;\rm sec$, which allows for an arbitrary initial time offset of the EMRI-disk system to match the observed timings, necessitated because the three initial phases of the Kerr geodesics were set to zero; and $\phi_{\rm disk,0}\sim[0,2\pi]$, which allows for an arbitrary initial phase offset between the accretion disk and EMRI orbit. We assumed a fixed $M_\bullet=10^{5.8}M_\odot$ \citep{Wevers22}, as in this exercise we are not interested in independently inferring the MBH mass, and making some choice is necessary to break the $a-M_\bullet$ degeneracy. We also fix $\chi_\bullet=0.5$ and $\theta_{\rm obs}=\pi/2$, because they have little to no effect on the timings for these large orbital radii and short baselines. In total, the free parameters of this fit are \{$a$, $e$, $i$, $\theta_{\rm disk}$, $T_{\rm disk}$, $\phi_{\rm disk, 0}$, $t_0$\}.

We find the posteriors are extremely multimodal, so we do not report the recovered parameters for further astrophysical inference. In Fig.~\ref{fig:ero1} we show the predicted orbiter-disk collision timings compared to the real data, showcasing that as a proof-of-concept, rigid disk precession appears able to roughly capture the observed erratic burst timings for reasonable parameter choices. As discussed in Section~\ref{subsec:with_precess}, sources such as these make inferring $M_\bullet$ and $a$ difficult, and we find even $e$, $i$, $\theta_{\rm disk}$ and $P_{\rm disk}$ are not uniquely constrained here. 

The fact that no single solution can fit all epochs together is almost certainly indicative of secular evolution in the EMRI-disk system, e.g. due to orbital decay, disk alignment, and changes in the disk density profile or precession speed. As we have not accounted for these effects in our short-timescale model, it cannot fully fit all epochs coherently in its current form. However, as QPE monitoring baselines grow longer, including secular effects in timing models is well-motivated, and will be the subject of future work (see \citealt{Zhou2024c,Xian2025} for further discussion).

\section{Discussion} \label{sec:discussion}
QPEs provide a novel window into orbital dynamics in galactic nuclei. If powered by repeated interactions between the accretion disk of an MBH and an orbiting secondary, their timing encodes key orbital and system parameters, including the spin and mass of the central black hole, and the eccentricity and semi-major axis of the secondary. In Section~\ref{sec:results} we have explored the constraining power of QPE timings alone in parameter inference of the MBH and EMRI properties. Two noteworthy shortcomings of our methods are:
\begin{enumerate}
    \item MCMC-based inference, which is notoriously susceptible to stalling at local minima for multimodal posterior surfaces (e.g. Fig.~\ref{fig:corner_withprecess}). As a result, it is possible the errors we report are over-estimates of the best-case constraints that may be obtained by more sophisticated sampling methods such as nested sampling (see e.g. \citealt{Zhou2024c} for comparison).
    \item Short observational baselines consisting of $\mathcal{O}(10-100\rm s)$ of QPEs, over which longer-term effects such as orbital decay and EMRI nodal precession driven by frame-dragging around a spinning MBH, do not set in. As a result, we do not produce any constraints on $\chi_\bullet$ and do not address prospects for measuring $\dot{P}$, which have also been addressed in other studies \citep{Xian2021,Zhou2024c}.
\end{enumerate}
Nevertheless, our work shows that, at least for some EMRI orbital parameters and disk properties, QPE timing may be used to infer $a$, $e$, $M_\bullet$, $\theta_{\rm disk}$, and $T_{\rm disk}$, to varying degrees of precision. Once inferred, these parameters become powerful tools for probing a range of astrophysical questions: How do massive black holes grow? What seeds them? Which channels form EMRIs? And how do QPEs couple to their accretion environments? In this section, we discuss what astrophysics becomes accessible through inference on each parameter.

\subsection{What can we learn from $M_\bullet$?}
Constraining the mass of the central black hole in QPE-hosting systems offers a unique opportunity to probe the less well-understood low-mass end of the black hole mass function (BHMF), defined as
\begin{equation}
\Phi(M_{\bullet}, z) = \frac{dN}{d\log M_{\bullet}\, dV},
\end{equation}
which quantifies the comoving number density of black holes per logarithmic mass interval at a given redshift. While the high-mass end of the BHMF is relatively well constrained through stellar and gas dynamical measurements in massive galaxies \citep{McConnell2013} and reverberation mapping of luminous AGN \citep{Peterson2004}, the low-mass end (\(M_{\bullet} \lesssim 10^6\,M_{\odot}\)) remains uncertain due to observational selection effects and the difficulty of detecting and characterizing faint or inactive nuclei in low-mass galaxies \citep{Greene2007}. 

In this regime, even the black hole occupation fraction is not well known, and different MBH seeding models (e.g., light vs. heavy seeds) make divergent predictions. Light seed models predict a steep low-mass slope and high occupation fraction in dwarf galaxies, while heavy seed models yield a suppressed BHMF below $10^5-10^6 M_{\odot}$  and a low occupation fraction in low-mass hosts \citep{Volonteri2010,Greene2020,Ricarte2018}. QPEs provide a new dynamical channel to infer MBH masses in otherwise quiescent or low-luminosity systems, potentially enabling direct constraints on the local BHMF and the demographics of the smallest massive black holes. Thus far, TDEs have been the dominant observational probe of low-mass MBHs; QPE timing may provide a promising method for refining the mass measurements in this population (see also \citealt{Zhou2025}), but as shown in Section~\ref{sec:results}, large numbers of bursts may be required for reliable constraints.

\subsection{What can we learn from $e$ and $a$?}
The distribution of orbital semimajor axes ($a$) and eccentricities ($e$) of QPE-hosting EMRIs offer constraints on their formation channels, which is relevant for estimating the EMRI rate observable by \textit{LISA}. EMRIs formed via the dry loss-cone channel or the Hills mechanism are expected to be highly eccentric (\(e \gtrsim 0.9\)), as they are injected onto plunging orbits through dynamical scatterings or binary disruptions in the nuclear stellar cluster \citep{Hopman2005,Preto2010,Broggi2022,Miller2005,Raveh2021}. In contrast, EMRIs formed in the wet channel, via migration through an accretion disc, experience efficient eccentricity damping due to interaction with the gaseous environment, yielding nearly circular orbits with \(e \sim h \ll 1\), where \(h\) is the local disc aspect ratio \citep{Lyu2025,Pan2021,Pan2021_3}. EMRIs formed via the Hills mechanism could retain high eccentricities even after extended GW-driven evolution (see e.g., Fig.\ 4 of \cite{Raveh2021}). Therefore, measurements of eccentricity from QPE-hosting systems can help provide evidence in favor of a wet or dry formation scenario.

The semi-major axis distribution is sensitively dependent on the orbital decay mechanisms. The GW-driven inspiral timescale scales steeply with the semi-major axis, approximately as \( t_{\rm GW} \propto a^4 \) for high-eccentricity systems \citep{Peters1964}, allowing predictions for when the secondary will either merge with the central MBH or be tidally disrupted. Moreover, since the companion crosses the accretion disc at a radius \( R_{\rm cross} \sim a(1 - e) \), the local gas density \( \rho(R) \propto R^{-s} \) and relative velocity \( v_{\rm rel} \sim \sqrt{G M_1 / R_{\rm cross}} \) directly influence the energy deposited during each crossing, and therefore the amount of orbital energy viscously dissipated to accelerate the inspiral.

\subsection{What can we learn from $\theta_{\rm disk}$ and $T_{\rm disk}$?}

The properties of accretion disks hosted in QPE systems---and more broadly, those around massive black holes---are a rich area of active study, as there remain longstanding uncertainties in their formation, stability, and structural properties \citep{Kara2025}. Most theoretical treatments of QPE accretion disks thus far assume either a standard steady-state $\alpha$-disk model \citep{Linial2023b,Yao2025a,Vurm2025,Lam2025} or do not explicitly model the disk \citep{Xian2021,Franchini2023}. Given the emergence of QPEs in the accretion flows formed by TDEs \citep{Chakraborty2021,Chakraborty2025a,Miniutti2023a,Quintin2023,Nicholl2024}, it is very likely the disks in these systems diverge significantly from this steady-state assumption. Recent work has explicitly taken steps toward treating the disk properties with the finite mass/energy/angular momentum reservoirs of compact TDE disks \citep{Guolo2025b,Mummery2025} by fitting the disk SEDs to infer the surface density profiles and radial extents. They indeed appear compatible with compact TDE disks extending to the EMRI orbital radii \citep{Mummery2020,Mummery2024b,Nicholl2024,Chakraborty2025a,Wevers2025,Guolo2025a}, adding mild circumstantial evidence to the EMRI-disk collision paradigm, albeit with some standing questions. Preliminary time-resolved studies of QPE X-ray spectra are also beginning to explore the accretion disk environment \citep{Kosec2025,Chakraborty2025b}, probing the properties of possible ionized outflows ejected from the disk in the QPE-generating process.

QPE timings probe orthogonal properties to the SED/spectral models, via the inclination angle and precession period. It has long been suggested that Lense-Thirring precession should be an observable phenomenon in TDE disks---e.g.\ via quasi-periodic oscillations of the disk emission---provided the accretion rate is high enough that the precession period exceeds the sound-crossing time in the disk. This is equivalent to requiring thick-disk conditions, i.e. $H/R\gtrsim \alpha$, where $\alpha$ is the viscosity parameter, $H$ the scale height, and $R$ the precession radius \citep{Stone2012,Franchini2016}. As the disk precession frequency depends jointly on the MBH spin, mass, and disk density profile, detection of precession and progressive alignment could also constrain the MBH properties. QPEs thus offer the possibility of studying the newly-formed, rapidly-evolving compact disks around MBHs via their dynamical imprints on the collision timings, adding a new method alongside the standard spectral fitting approach.

\subsection{Comparisons with prior work} \label{subsec:priorwork}

During the preparation of this work, two studies were posted \citep{Zhou2024c,Zhou2025} which explore a similar Bayesian inference paradigm for QPE timings with a rigidly precessing accretion disk. Our work was carried out independently, using different methods including the post-Newtonian formalism rather than exact Kerr geodesic trajectories, MCMC rather than nested sampling, and GPU acceleration, with the primary goal of computational efficiency. In particular, the costs of nested sampling and exact calculation of the Kerr geodesics are prohibitively expensive in many cases, requiring up to 640 CPU cores over 20 days ($\gtrsim 300,000$ CPU-hours) for a full inference run including the effects of disk precession and orbital decay \citep{Zhou2024c}. While we have not yet introduced orbital evolution into our code---and also therefore limited our focus to shorter observational baselines than considered in those works---our work provides a first step toward developing more inexpensive inference tools. 

The speed of our inference tools allowed us, for the first time, to systematically explore the entire EMRI/disk parameter space broadly. The QPE timing problem warrants independent investigations with a range of different approaches, particularly as the observational data grows and its relevance to the broader astrophysical community becomes more compelling.

\section{Conclusion} \label{sec:conclusion}

We have developed \qpefit, an efficient model to compute the timings of orbiter-disk collisions for a low-mass companion around a massive black hole (MBH) hosting an accretion disk. The model is relevant for the recently-identified Quasi-Periodic Eruptions (QPEs), which have been proposed to be powered by star- or BH-disk collisions in an extreme mass-ratio inspiral (EMRI). In principle, the burst timings encode information about the EMRI orbital trajectories, including the general relativistic precessions affecting both the orbiter and the accretion disk; thus, precise measurement of QPE timings may allow inference of the properties of the MBH (mass, spin), EMRI orbit (semi-major axis, eccentricity), and accretion disk (inclination angle, nodal precession timescale).

\qpefit\ computes the geodesic trajectory of the companion in the post-Newtonian formalism up to 3PN order, following the eccentricity expansion of \cite{Sago2015}. These analytic approximations, along with the significant acceleration afforded by GPUs (Table~\ref{tab:benchmark}), enable rapid computation of the trajectories, making them computationally amenable to Bayesian parameter inference. By carrying out an extensive grid of MCMC parameter inference simulations across a wide grid of injected EMRI/disk/MBH properties, we have quantified the characteristic errors on each inferred parameter. The results are presented and described in Section~\ref{sec:results}, Figures~\ref{fig:vary_noprecess}-\ref{fig:vary_obs}, and Table~\ref{tab:errors}. We find the following:
\begin{itemize}
    \item For the simple-case model of an EMRI colliding with a static, non-precessing accretion disk, QPE timings provide $\lesssim 10$\% error-bars on EMRI semimajor axis/eccentricity and MBH mass within tens of orbital periods (Fig.~\ref{fig:corner_noprecess}). The best constraints are for EMRIs with modest eccentricities $0.1\lesssim e\lesssim 0.5$ (Fig.~\ref{fig:vary_noprecess}).
    \item However, given that observed QPE timings appear more complicated than this simple picture, it is important to consider further effects influencing the timing, e.g rigid-body precession of the accretion disk (Fig.~\ref{fig:oc}). Introducing a precessing disk complicates parameter inference considerably, most significantly degrading MBH mass constraints (Fig.~\ref{fig:vary_withprecess}), although the disk properties themselves (e.g. inclination, precession rate) can be constrained to the $\sim 10-50\%$ level (Fig.~\ref{fig:corner_withprecess}, Table~\ref{tab:errors}). Performing inference with the static-disk model when the true underlying process involves a precessing disk can systematically induce biases in the inferred parameters (Fig.~\ref{fig:precessionbias}).
    \item Better constraints typically arise from monitoring campaigns with shorter observational gaps (Fig.~\ref{fig:vary_obs}), motivating high-cadence observations for constraints on MBH mass/EMRI properties.
    \item To showcase the ability of \qpefit\ to describe real QPE timing data, we applied it to the irregular timings of eRO-QPE1 (Fig.~\ref{fig:ero1}), finding reasonable fits but with time-varying parameters. This motivates self-consistently modeling the effects of secular period/eccentricity/disk evolution alongside the EMRI trajectory calculation.
\end{itemize}

In this paper, we only explored relatively short observational baselines of $\mathcal{O}(10-100\rm s)$ of QPEs, meaning we have \textit{not} considered timing effects from the secular evolution of the EMRI-disk system. These baselines also prohibit inference of MBH spin, as the effect of general relativistic frame-dragging only sets in on timescales $\gg P_{\rm orb}$. Both of these effects are inevitable consequences of the EMRI model, which predicts orbital decay driven by a combination of gravitational waves, gas drag, and/or dynamical friction \citep{Kejriwal2024,Linial2024a,Zhou2024c,Xian2025}. Long-term monitoring observations of some sources have even tentatively detected orbital period evolution in some sources \citep{Arcodia2024b,Zhou2024c,Miniutti2025,Xian2025}, though not yet with unambiguous solutions. Continued monitoring of the known QPEs is sure to place further constraints on models, and in future work, we will add secular evolution to the model presented here. Moreover, the posterior error values on some parameters are likely to be overestimated due to the inherent shortcomings of Markov Chain Monte Carlo (MCMC) inference, particularly for the multi-modal parameter spaces relevant here. Exploring more sophisticated inference schemes is thus well-motivated, and may provide significant improvements in parameter inference from QPE timings.

\section*{Acknowledgements}
We thank Katerina Chatziioannou and Ethan Payne for useful discussions. We also thank the anonymous referee for their helpful comments. The authors acknowledge the MIT Office of Research Computing and Data for providing high performance computing resources that have contributed to the research results reported within this paper. Additionally, some of the calculations in this work were conducted in the Resnick High Performance Computing Center, a facility supported by Resnick Sustainability Institute at the California Institute of Technology. LVD is supported by the Sherman Fairchild Postdoctoral Fellowship at the California Institute of Technology. SAH is supported by NSF Grants PHY-2110384 and PHY-2409644; NSF Grant PHY-2110384 also supported LVD at MIT during a portion of this work. GM thanks the Spanish MICIU/AEI/10.13039/501100011033 and ERDF/EU grants n.\ PID2020-115325GB-C31 and n.\ PID2023-147338NB-C21 for support. MG is funded by Spanish MICIU/AEI/10.13039/501100011033 and ERDF/EU grant PID2023-147338NB-C21. MB acknowledges support from the Italian Ministry for Universities and Research (MUR) program “Dipartimenti di Eccellenza 2023-2027”, within the framework of the activities of the Centro Bicocca di Cosmologia Quantitativa (BiCoQ). This work makes use of the Black Hole Perturbation Toolkit \citep{BHPToolkit}, in particular the KerrGeoPy package \citep{kerrgeopy}.

\bibliography{refs}{}
\bibliographystyle{aasjournal}

\appendix
\counterwithin{figure}{section}
\counterwithin{table}{section}

\FloatBarrier
\section{Impact of the Post-Newtonian approximation in the  trajectory } \label{subsec:pn_vs_kerr}

\begin{figure}
    \centering
    \includegraphics[width=\textwidth]{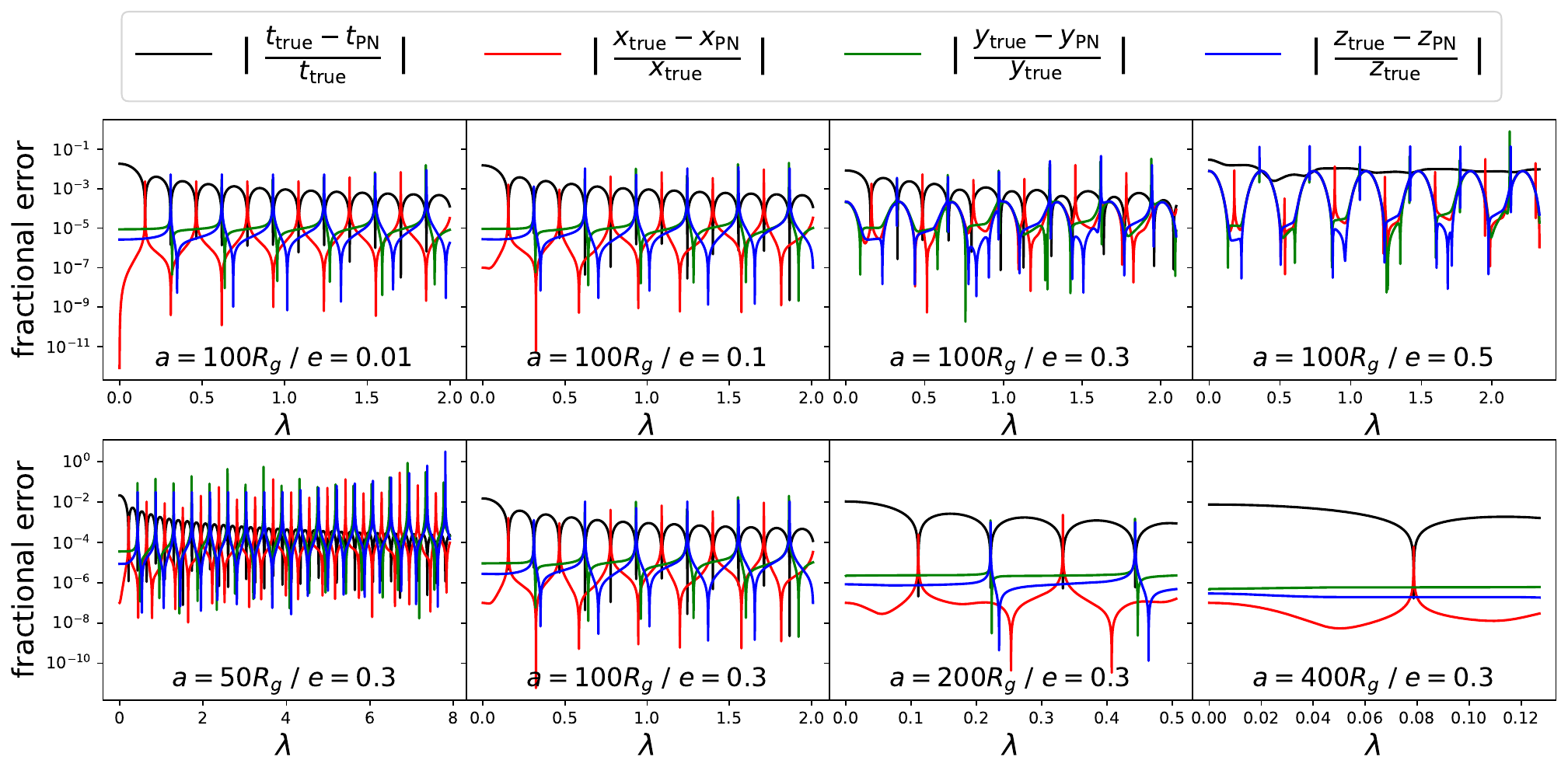}
    \caption{Deviation of 3PN trajectories from the exact Kerr geodesics as a function of Mino time ($\lambda$). All trajectories were computed with 10 second time resolution over a 100,000 second baseline.}
    \label{fig:pn_vs_kerr}
\end{figure}

Figure~\ref{fig:pn_vs_kerr} quantifies the deviation of 3PN orbital trajectories from exact Kerr geodesics for a range of orbital configurations. The deviation increases noticeably above eccentricities of 0.5, demonstrating the reduced accuracy of the PN model in describing highly eccentric motion. At fixed $e=0.3$, we also observe that deviation generally decreases with increasing $a$, consistent with improved PN convergence in the weak-field regime. The PN model performs best for low-eccentricity, large-radius orbits.

\begin{figure}
    \centering
    \includegraphics[width=0.65\textwidth]{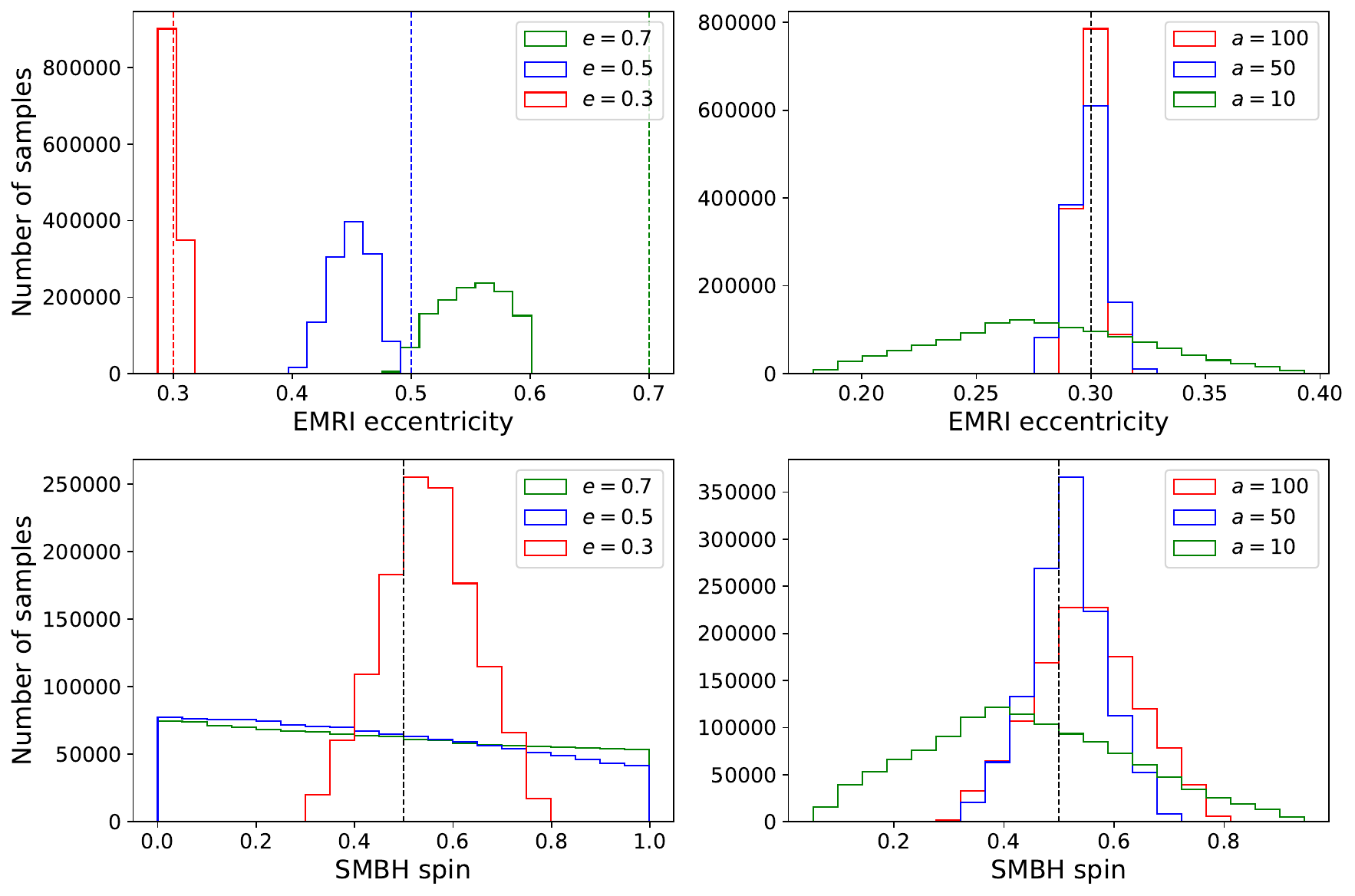}
    \caption{Recovered distributions of eccentricity ($e$) and MBH spin ($\chi_\bullet$) for different true values, demonstrating the regime of applicability of the PN approximation for parameter recovery.
  \textbf{Top left panel:} Recovered $e$ for true $e=0.7$ (green), $0.5$ (blue), and $0.3$ (red). 
  \textbf{Top right panel:} Recovered $e$ for true $a=100\,R_g$ (red), $50\,R_g$ (blue), and $10\,R_g$ (green). 
  \textbf{Bottom left panel:} Recovered $\chi_\bullet$ for different values of $e$ (same color scheme as top-left). 
  \textbf{Bottom right panel:} Recovered $\chi_\bullet$ for different values of $a$(same color scheme as top-right). 
  Dashed vertical lines mark the true values.}
    \label{fig:ecc_a_recovery}
\end{figure}

To examine the range over which the Post-Newtonian (PN) timing model remains accurate for parameter inference, we inject synthetic timing data using exact KerrGeoPy trajectories and then recover the eccentricity ($e$) and MBH spin ($\chi_\bullet$) with a 2D PN-based likelihood.  All other EMRI parameters (mass, inclination, etc.) are held fixed at their true values, such that any deviations in the posterior directly reflect PN truncation errors in the recovered parameters. Figure~\ref{fig:ecc_a_recovery} presents the resulting posterior distributions.  The two panels on the left show recovered $e$ and $\chi_\bullet$ for three injected eccentricities ($e = 0.3,\,0.5,\,0.7$ at fixed $a=100\,R_g$), and the panels on the right show the same recovered quantities for three injected semi-major axes ($a = 10,\,50,\,100\,R_g$ at fixed $e=0.3$).

In the top-left panel, both the width of the posterior distributions and the bias in the recovered eccentricities increase with the true value of $e$. At $e=0.5$, which marks the formal limit of post-Newtonian (PN) applicability, the posterior is noticeably broader and shifted, indicating a clear bias in the recovered eccentricity. At $e=0.7$, the PN expansion fails, resulting in a breakdown of the recovery. In the bottom-left panel, which shows the recovered semi-major axis ($a$) as a function of $e$, accurate recovery of $e$ (e.g., at $e=0.3$) leads to well-constrained posteriors on $\chi_\bullet$. However, for $e=0.5$ and $e=0.7$, where the eccentricity estimates are biased or highly uncertain, the posterior on $a$ becomes extremely broad and effectively unconstrained.

In the bottom right panel where the $\chi_\bullet$ value is recovered, spin inference shows a clear dependence on the injection radius. At large radii ($a = 100\,R_g$), the timing signal becomes less sensitive to the effects of $\chi_\bullet$. As a result, the posterior widens and its median shifts modestly. At small radii ($a = 10\,R_g$), deep in the strong-field regime, spin effects such as frame dragging are enhanced, but PN truncation errors dominate, leading to a broad and significantly biased posterior. In contrast, at $a = 50\,R_g$, the PN model remains valid and the timing remains sufficiently sensitive to spin, yielding the most accurate and tightly constrained $\chi_\bullet$ posterior. This trend reflects competing influences: diminished spin sensitivity in the weak field versus poor PN convergence in the strong field, with $a \sim 50\,R_g$ emerging as an optimal regime for PN-based spin recovery. In the recovered-$e$ panels for varying $a$, eccentricity is accurately recovered at $a = 100\,R_g$, with modest broadening at $a = 50\,R_g$ as the PN approximation begins to deteriorate. At $a = 10\,R_g$, the posterior becomes broad and biased, consistent with failure of the PN model. Unlike spin, the influence of eccentricity is not strongly relativistic and can still be inferred reasonably well in the weak-field, provided the PN series remains convergent.

These 2D posteriors define practical boundaries for PN-based inference; the PN timing model reliably recovers $(\chi_\bullet,e)$ for $e\lesssim0.4$ at $a\gtrsim50\,R_g$, and should be replaced by a fully relativistic trajectory (e.g.\ KerrGeoPy \citep{kerrgeopy}) outside that regime. Note that because we performed a restricted two-parameter inference in this section, the posterior widths are inherently narrower, making biases appear more pronounced than they would in a full-parameter model.

\section{MCMC convergence}
\label{appendix:convergence}
\begin{figure}
\centering
    \includegraphics[width=0.8\linewidth]{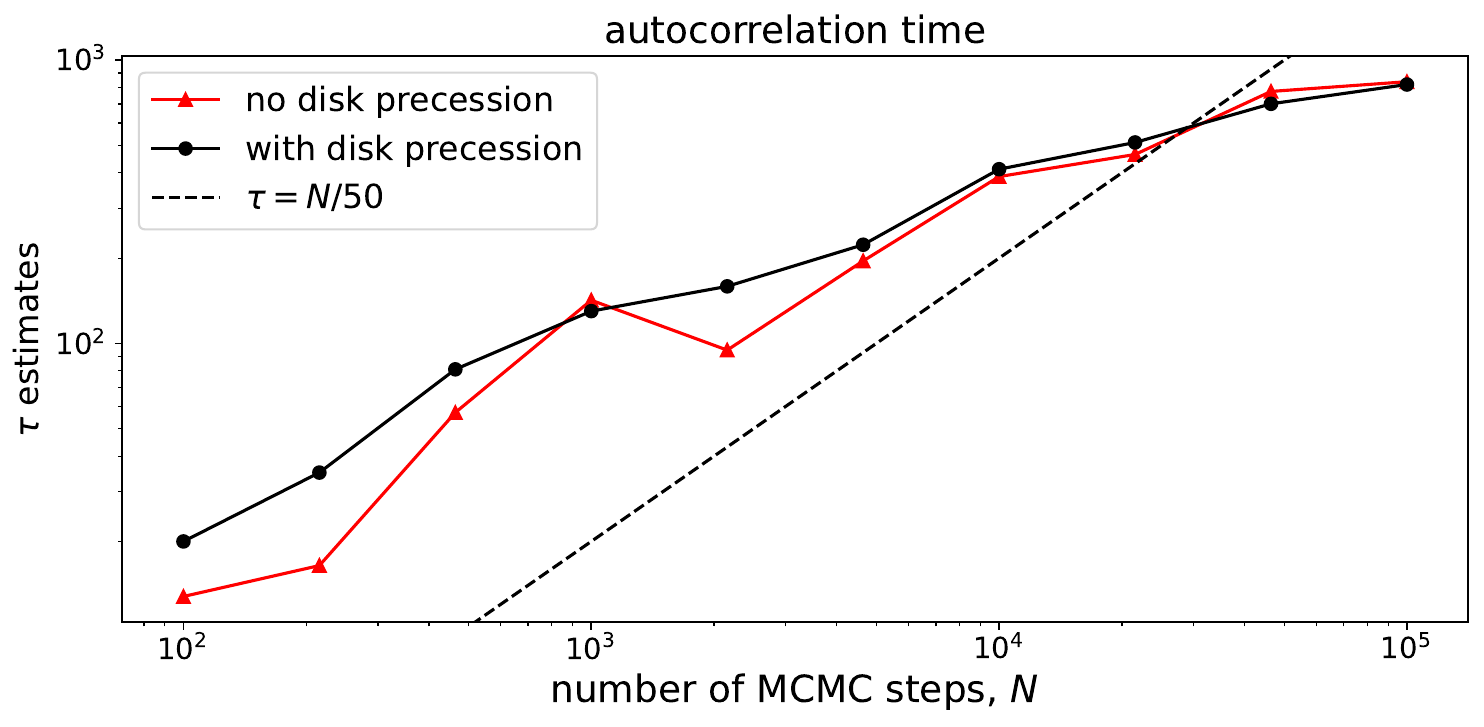}
    \includegraphics[width=0.8\linewidth]{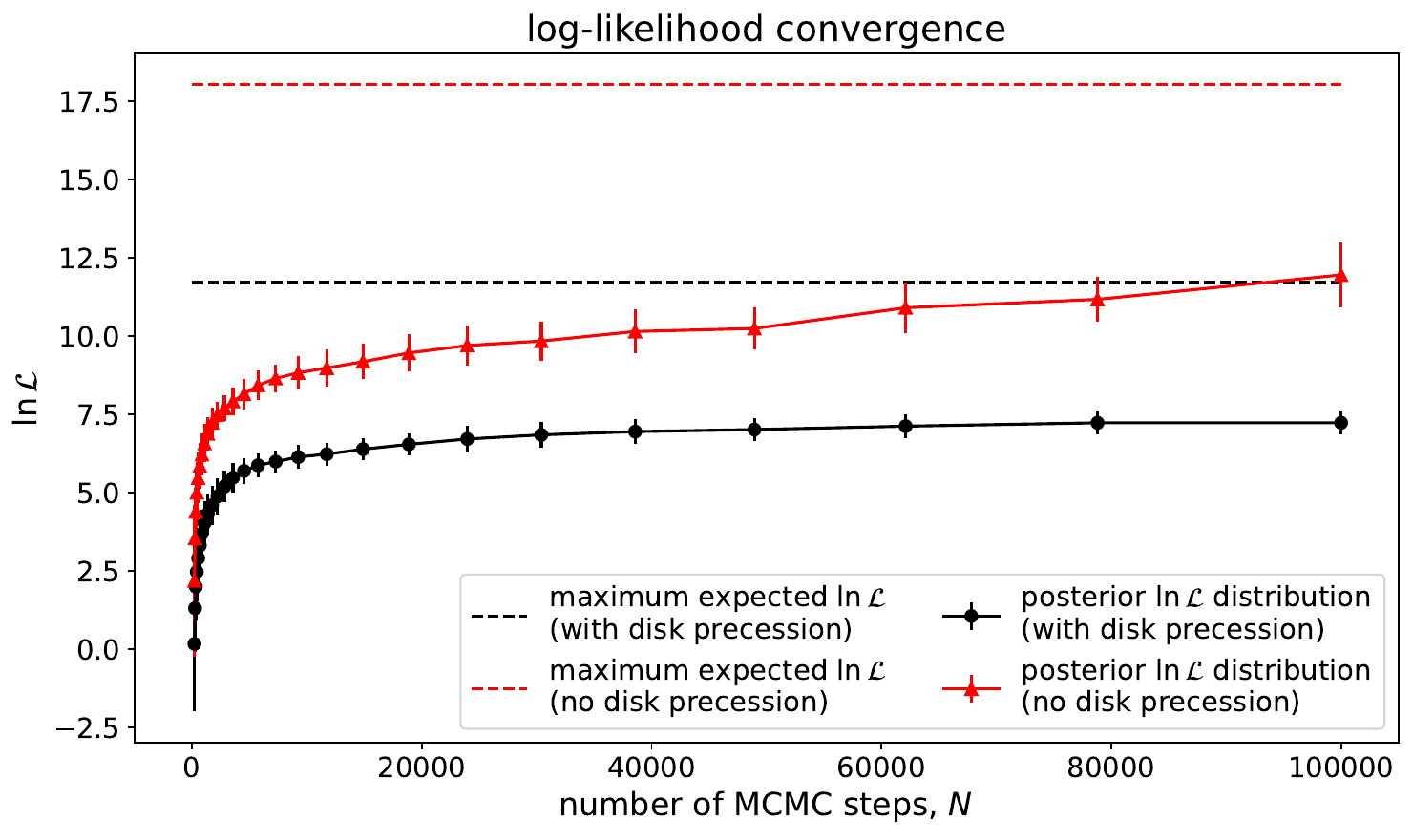}
    \caption{\textbf{Top panel:} integrated autocorrelation time of the MCMC walkers, estimated by the method of \cite{Goodman2010}. We show two illustrative cases, both using $a=100R_g$, $e=0.1$, $M_\bullet=10^5M_\odot$, and $\chi_\bullet=0.5$. The red triangles are for a sampling run with no disk precession, and the black circles introduce a misaligned precessing disk with $\theta_{\rm disk}=20^\circ$ and $T_{\rm disk}=20P_{\rm orb}$. \textbf{Bottom panel:} for the same sampling runs above, we show the mean and standard deviations of the posterior likelihood distribution, which are chosen by retaining the $10^5$ highest-likelihood samples from the chain.}
    \label{fig:convergence}
\end{figure}

In Fig.~\ref{fig:convergence} we show two illustrative metrics addressing the convergence of the MCMC chains, for two sampling runs presented in Figs.~\ref{fig:corner_noprecess} \&~\ref{fig:corner_withprecess} ($a=100R_g$, $e=0.1$, $M_\bullet=10^5M_\odot$, $\chi_\bullet=0.5$, and for the case with disk precession, $\theta_{\rm disk}=20^\circ$, $T_{\rm disk}=20P_{\rm orb}$). In the top panel, we show the integrated autocorrelation time \citep{Goodman2010}, which is a crude proxy for assessing whether the chains are properly converged. While there is no formally defined criteria for acceptable convergence, typically an autocorrelation time ($\tau$) smaller than $\tau \lesssim N/50$ (where $N$ is the number of MCMC steps) is taken as indication that the chain is reasonably converged (e.g. see the discussion in \href{https://emcee.readthedocs.io/en/stable/tutorials/autocorr}{https://emcee.readthedocs.io/en/stable/tutorials/autocorr/}). Our samplers reach this threshold roughly at 20,000 steps, and are well below it within $10^5$ steps, providing some indication that the results presented in Section~\ref{sec:results} are reasonable.

The bottom panel of Fig.~\ref{fig:convergence} shows an even clearer indication that running the chains longer will not produce significant additional returns. We show the mean and standard deviation of the posterior likelihood distributions pictured in Figs.~\ref{fig:corner_noprecess} \&~\ref{fig:corner_withprecess}, which plateau within about $10^4$ steps and increase only modestly thereafter. We compare this to the ``theoretical'' maximum expected likelihood, which is obtained by evaluating the likelihood function for the true injected parameters. The posterior likelihood distribution falls noticeably short of this maximum, indicating the inherent shortcoming of MCMC for exploring this likelihood space within a reasonable number of steps.

\FloatBarrier
\section{Impact of observing angle}
\label{sec:obsangle}
\begin{figure*}
    \centering
    \includegraphics[width=\linewidth]{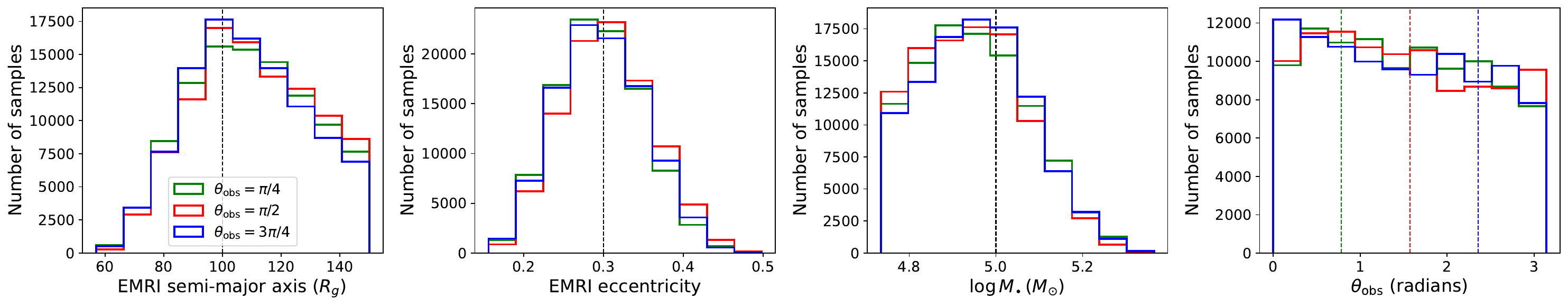}
    \caption{Parameter recovery for various $\theta_{\rm obs}$ values ($\pi/4$, $\pi/2$ and $3\pi/4$) for an EMRI with $a=100R_g$ and $e=0.3$ around an MBH with $M_\bullet=10^5M_\odot$, $\chi_\bullet=0.5$. We use 500 independent walkers, 5000 steps, 50\% burn-in and plot the top $10^5$ highest-likelihood samples.}
    \label{fig:bias_histograms_counts}
\end{figure*}

We examine the impact of varying the observing angle $\theta_{\rm obs}$ on parameter recovery. Table~\ref{tab:obsangle} summarizes the recovered MAP values and deviations from the true values for the semi-major axis, eccentricity, and black hole mass across $\theta_{\rm obs} = \pi/2$, $3\pi/4$, and $\pi/4$, while Figure~\ref{fig:bias_histograms_counts} shows the corresponding posterior distributions. We find that the inferred parameters are largely insensitive to the choice of $\theta_{\rm obs}$: deviations from the truth remain below $0.3\sigma$ in all cases. The semi-major axis exhibits the largest spread, with the MAP value decreasing from $103.00$ at $\theta_{\rm obs} =\pi/4$ to $95.38$ at $3\pi/4$, indicating a weak trend with observing angle. The greatest deviation in eccentricity occurs at $\theta_{\rm obs} = \pi/2$ (0.26$\sigma$), while the largest deviation in black hole mass appears at $3\pi/4$ (0.26$\sigma$). 

\begin{table*}
\centering
\caption{Recovered maximum a posteriori values and deviations from the true values for the semi-major axis, eccentricity, and black hole mass, comparing different observing inclination angles $\theta_{\rm obs}$.}
\label{tab:obsangle}
\begin{tabular}{lccccc}
\toprule
Parameter & $\theta_{\rm obs}=\pi/4$ & $\theta_{\rm obs}=\pi/2$ & $\theta_{\rm obs}=3\pi/4$  \\
\midrule
\midrule
Semi-major axis $a$ & 103 (0.15$\sigma$)  & 103 (0.14$\sigma$) & 95 (-0.23$\sigma$) \\
Eccentricity $e$ & 0.298 (-0.04$\sigma$)& 0.315 (0.26$\sigma$) & 0.300 (0.02$\sigma$)  \\
BH Mass & 4.98 (-0.15$\sigma$) & 4.98 (-0.14$\sigma$) & 5.03 (0.26$\sigma$)  \\
\bottomrule
\end{tabular}
\end{table*}

These trends are modest, and overall changes in $\theta_{\rm obs}$ introduce only minor variations in parameter recovery. We conclude that uncertainties related to the observing angle are subdominant compared to other sources of modeling bias, such as whether disk precession is properly included.

\section{Corner plots of data fits to eRO-QPE1} \label{appendix:ero1_corner}

\begin{figure}[htbp]
    \centering
    \begin{minipage}{0.42\textwidth}
        \centering
        Epoch 1 (Aug 2020)
        \includegraphics[width=\textwidth]{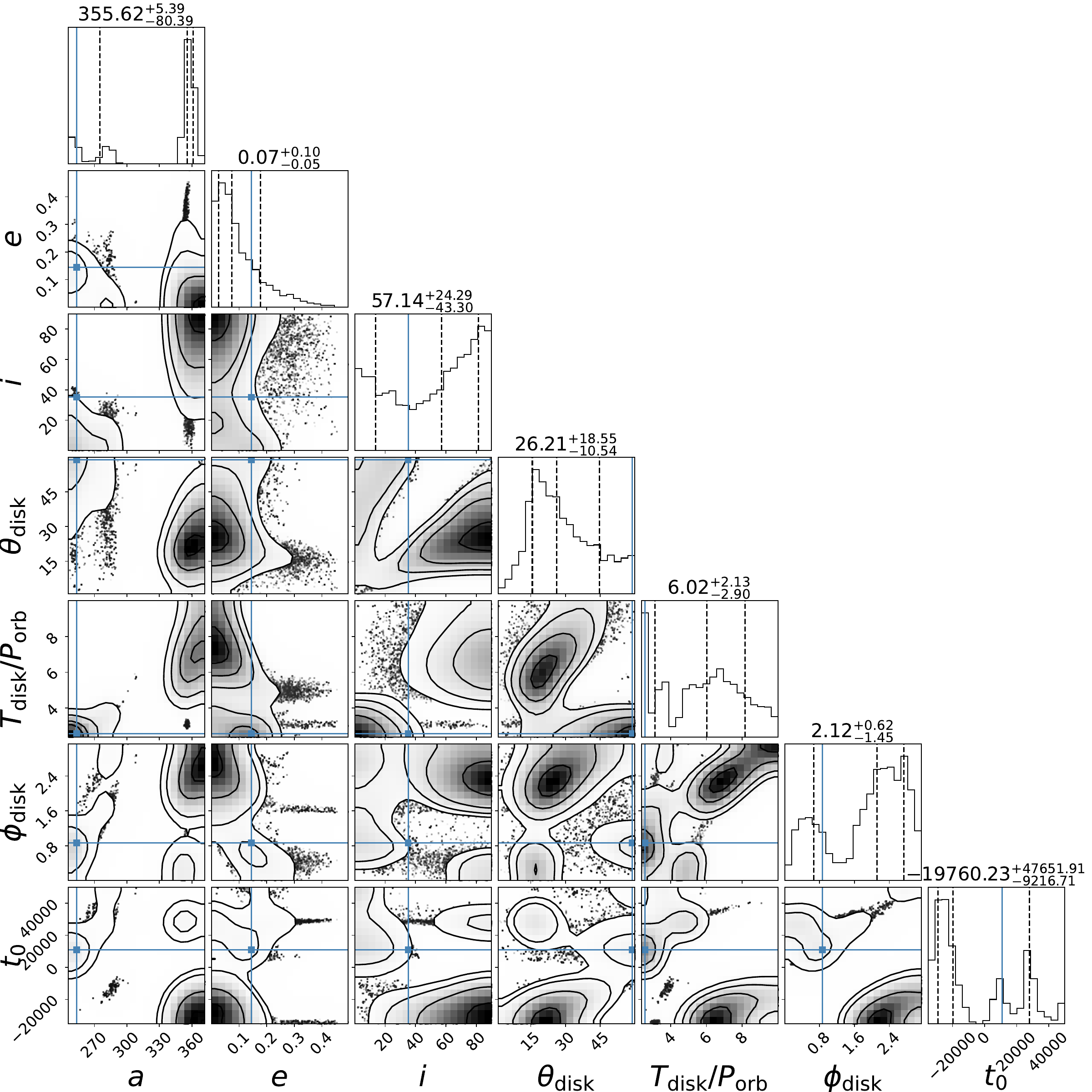}
    \end{minipage}
    \hfill
    \begin{minipage}{0.42\textwidth}
        \centering
        Epoch 2 (Aug 2021)
        \includegraphics[width=\textwidth]{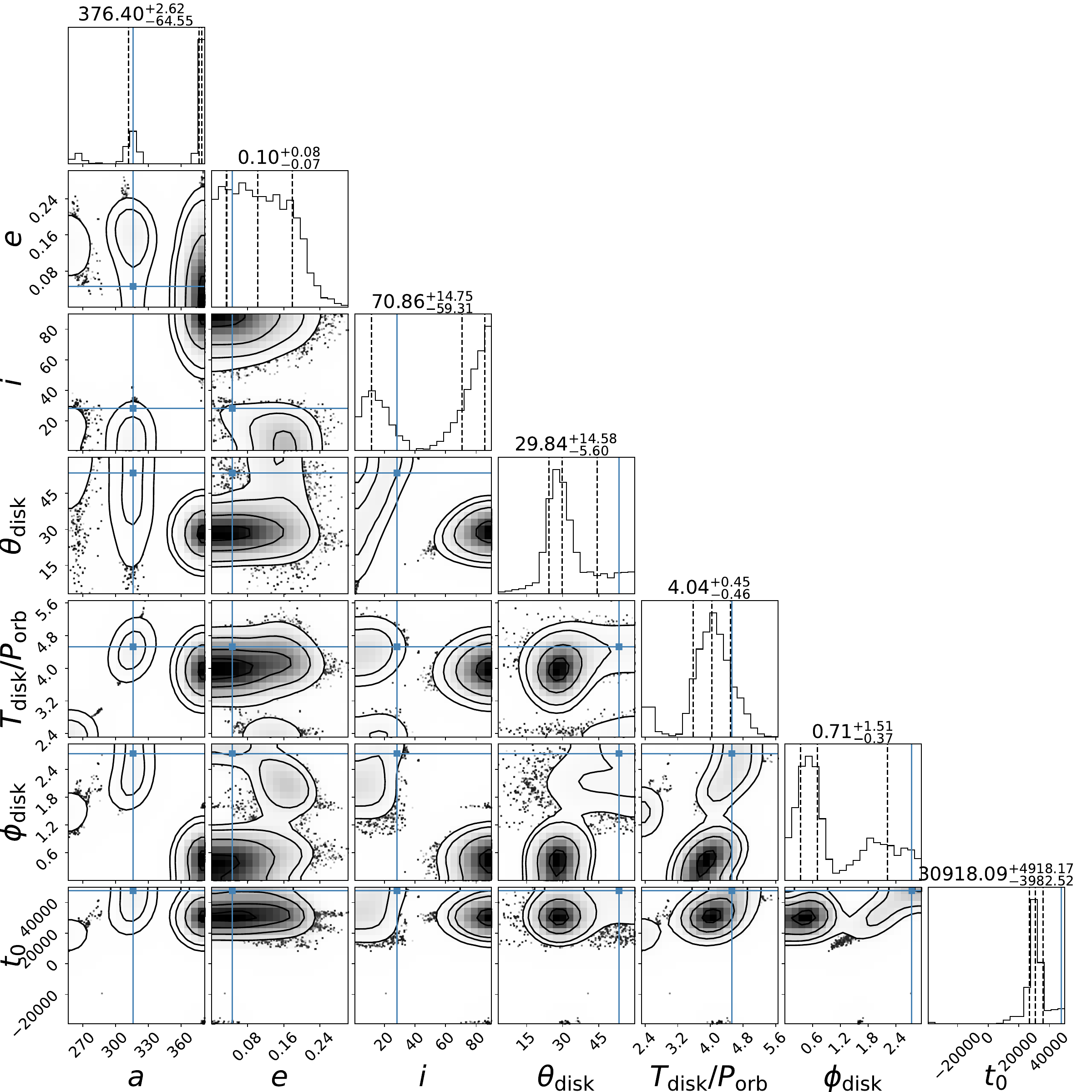}
    \end{minipage}
    \\[1em]
    \begin{minipage}{0.42\textwidth}
        \centering
        Epoch 3 (Feb 2022)
        \includegraphics[width=\textwidth]{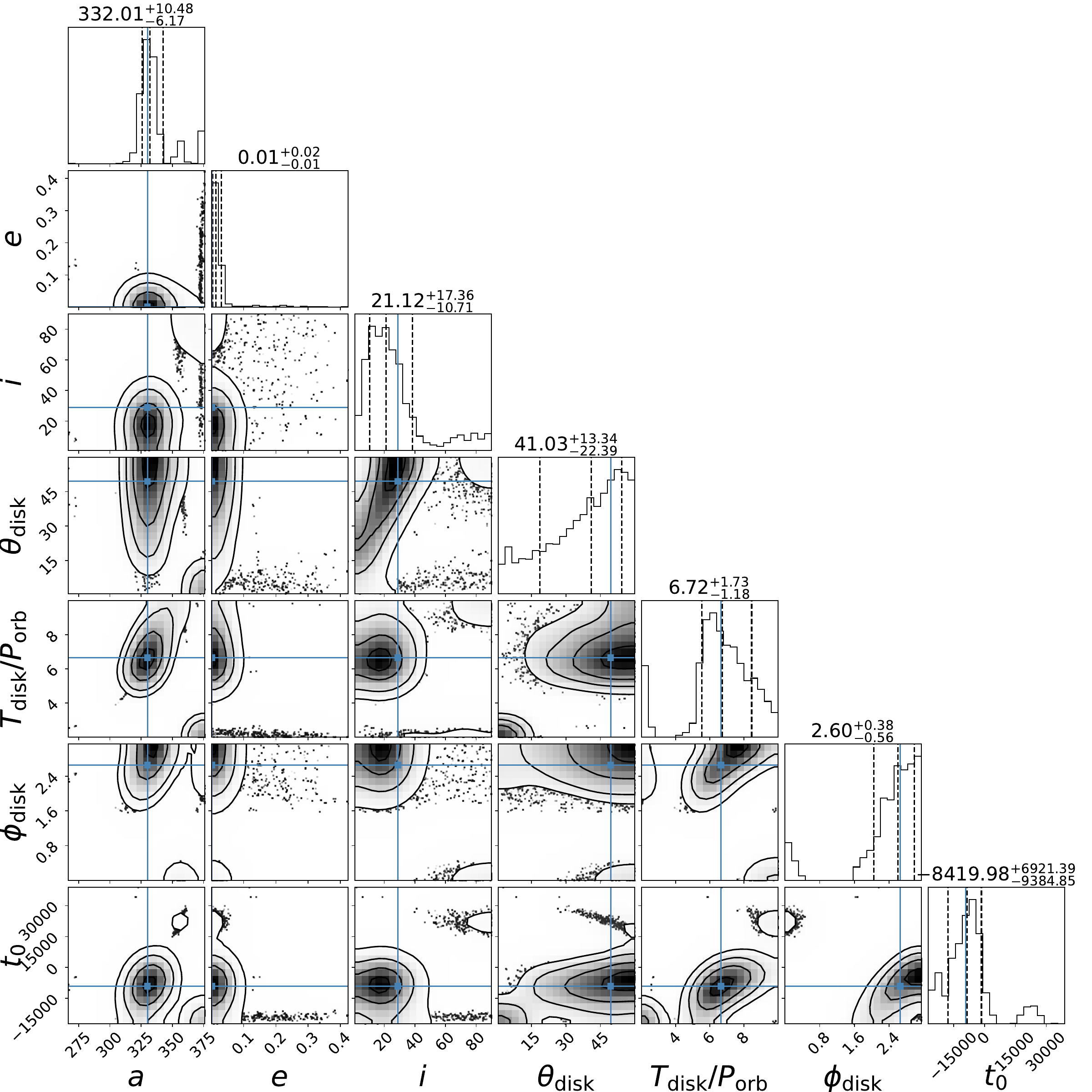}
    \end{minipage}
    \hfill
    \begin{minipage}{0.42\textwidth}
        \centering
        Epoch 4 (Dec 2022)
        \includegraphics[width=\textwidth]{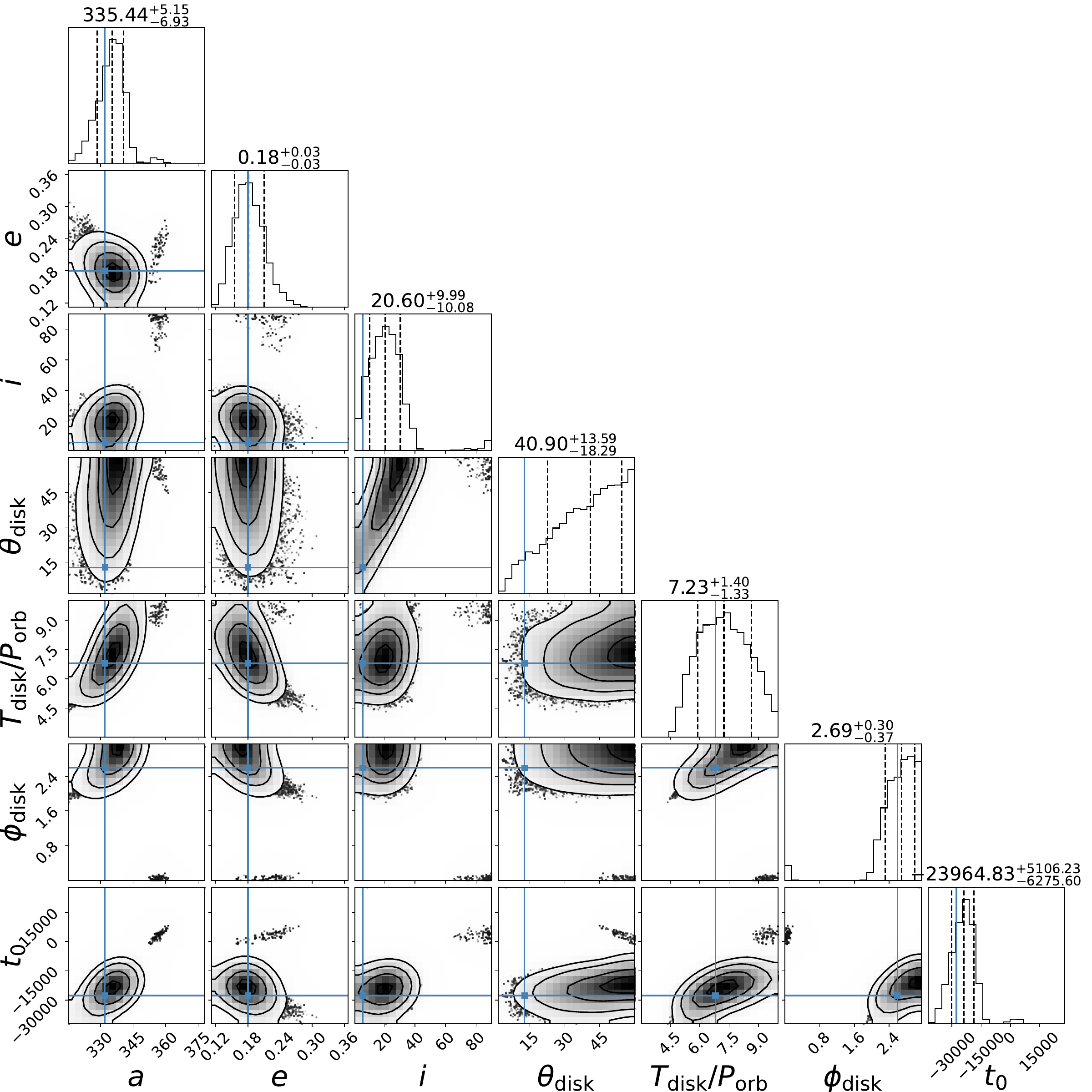}
    \end{minipage}
    \\[1em]
    \begin{minipage}{0.42\textwidth}
        \centering
        Epoch 5 (Jan 2023)
        \includegraphics[width=\textwidth]{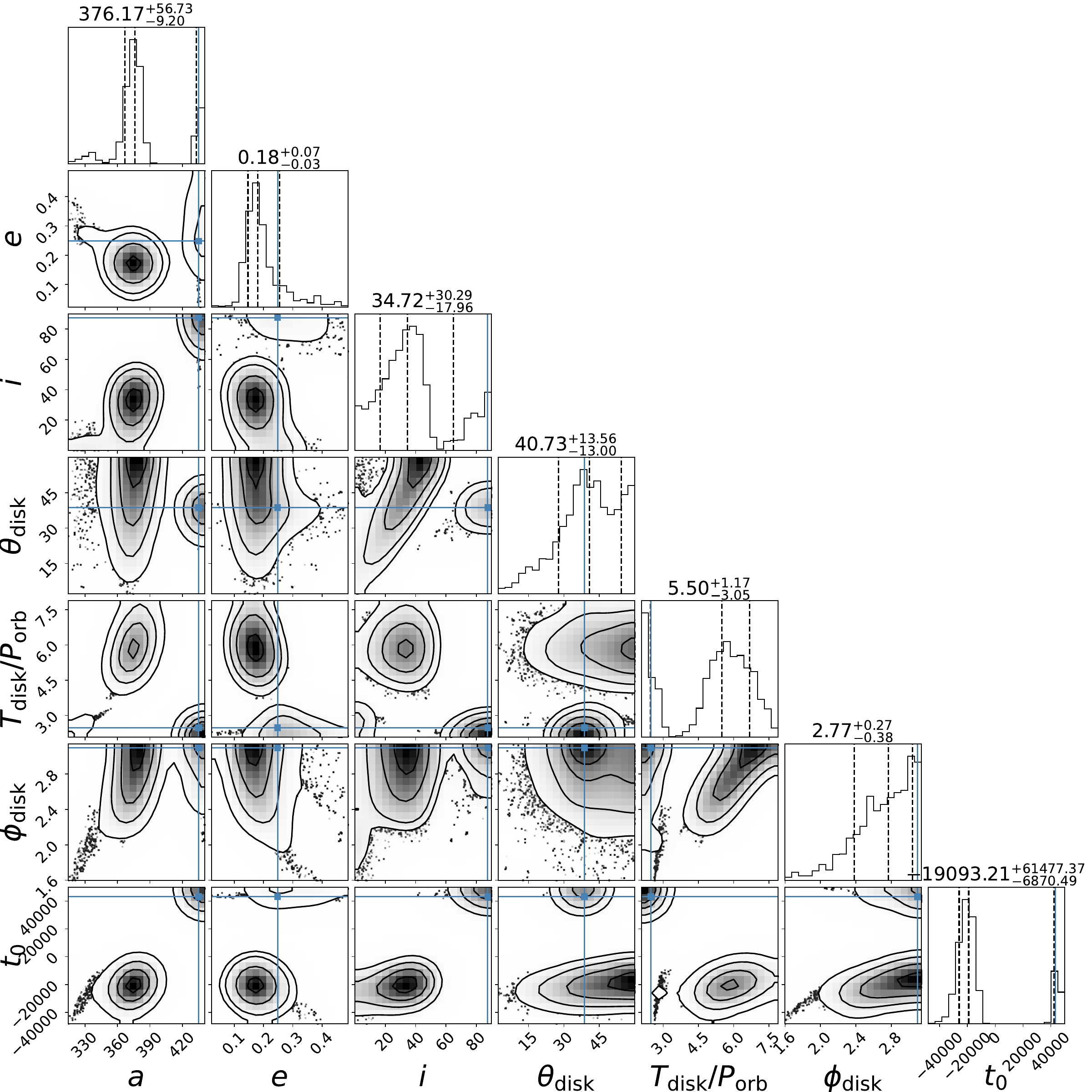}
    \end{minipage}
    \caption{Epoch-by-epoch corner plots for fits to eRO-QPE1 \textit{NICER} data.}
    \label{fig:ero1_corners}
\end{figure}

In Fig.~\ref{fig:ero1_corners}, we show corner plots of each of the individual-epoch fits to the \textit{NICER} data of eRO-QPE1, as outlined in Section~\label{subsec:application}. The posteriors are smoothed by a $2\sigma$ gaussian kernel. The blue horizontal/vertical lines indicate the locations of the maximum-likelihood parameter combination, which were used to produce the simulated timings presented in Fig.~\ref{fig:ero1}. In many cases, the maximum-likelihood parameters differ significantly from the mean/median of the posteriors, indicating a complex multimodality in the parameter space. While we have shown in principle that the EMRI+rigid disk precession model is capable of producing timings consistent with QPE observations, this suggests that significant caution must be taken when interpreting parameter estimates from QPE data.

\section{Results across all simulations} \label{appendix:all_sim}

To explore as wide a region of parameter space as computationally feasible, we ran a large grid of MCMC sampling runs for different EMRI/disk parameter combinations with and without disk precession included. For the static aligned-disk model, we sampled over every combination spanning $a\in\{100,200,400\}R_g$, $e\in\{0.01,0.1,0.3,0.5\}$, $\chi_\bullet\in\{0.5,0.9\}$, $\log (M_\bullet/M_\odot)\in\{5,6,7\}$, and observing strategies from \{$1\times100$ ks, $4\times 100$ ks with 200 ks gaps, and $4\times 100$ ks with 1 Ms gaps\}, for a total of $3\times 4\times 2\times 3\times 3=216$ combinations.  For the misaligned precessing disk model, we included all of the above parameters, plus $T_{\rm disk}=\{5,20,50\}P_{\rm orb}$ and $\theta_{\rm disk}=\{5^\circ,20^\circ\}$, for a total of $216\times 3\times 2 = 1296$ samples. All MCMC sampling was performed on NVIDIA A100 GPUs, using 1000 walkers $\times$ 20000 steps with a 10-second time resolution for the trajectory computation. The sampling runs with a static disk spanned between 1-1.5 hr, with an average duration of 70.5 minutes, while the sampling runs for a precessing disk spanned between 1.5-8 hr, with an average duration of 280.8 minutes. In total, this amounts to 6319 GPU-hours, which we carried out across 16 NVIDIA A100s. In Table~\ref{tab:errors} we report the average 16th/50th/84th percentile parameter recovery, averaged over all combinations, to show the typical error-bars expected from QPE timing inference. We used uniform priors of $a\sim \mathcal U(30, 500)R_g$, $e\sim \mathcal U(0, 0.9)$, $i \sim \mathcal U(0^\circ, 90^\circ)$, $\chi_\bullet \sim \mathcal U(0, 0.998)$, $\log (M_\bullet/M_\odot) \sim \mathcal U(4, 8)$, $\theta_{\rm disk} \sim \mathcal U(0^\circ, 60^\circ)$, and $T_{\rm disk} \sim \mathcal U(2 , 100)P_{\rm orb}$. The restrictions of $i \sim \mathcal U(0^\circ,90^\circ)$ and $\theta_{\rm disk}\sim\mathcal U(0^\circ, 60^\circ)$ are made for simplicity, but they exclude regions of parameter space which are formally possible---retrograde EMRIs and highly-inclined, rigidly precessing disks. We note, however, that EMRI inclination has minimal effect on QPE timings (Section~\ref{sec:results}), and it is furthermore unclear whether extremely inclined disks can undergo rigid precession as unmodeled non-linear effects such as disk tearing \citep{Nixon2012,Kaaz2023} become potentially important.

\begin{longtable*}{lc|cc|cc}
\caption{Average 16/50/84th percentile of posteriors, marginalized over other parameters, for our large grid of simulations.}
\label{tab:errors}
\endfirsthead
\toprule
\toprule
 & 
  & \multicolumn{2}{c|}{Static aligned disk} & \multicolumn{2}{c}{Misaligned precessing disk} \\
\cmidrule(lr){3-4} \cmidrule(lr){5-6}
Parameter & Observing strategy & True val. & Average est. & True val. & Average est. \\
\midrule
\multirow{9}{*}{EMRI semimajor axis ($R_g$)} 
  & \multirow{3}{*}{$1\times100$~ks} & 100 & $114^{+16}_{-18}$ 
                                     & 100 & $178^{+91}_{-73}$ \\
  &                                  & 200 & $211^{+23}_{-29}$ 
                                     & 200 & $312^{+100}_{-84}$ \\
  &                                  & 400 & $405^{+30}_{-35}$
                                     & 400 & $354^{+98}_{-122}$ \\
\cmidrule(lr){2-6}
  & \multirow{3}{*}{$4\times100$~ks (200~ks gaps)} & 100 & $108^{+17}_{-16}$ 
                                                   & 100 & $187^{+102}_{-80}$ \\
  &                                                & 200 & $209^{+21}_{-23}$ 
                                                   & 200 & $268^{+99}_{-86}$ \\
  &                                                & 400 & $402^{+25}_{-25}$ 
                                                   & 400 & $342^{+94}_{-107}$ \\
\cmidrule(lr){2-6}
  & \multirow{3}{*}{$4\times100$~ks (1~Ms gaps)} & 100 & $110^{+16}_{-20}$ 
                                                 & 100 & $227^{+123}_{-92}$ \\
  &                                              & 200 & $206^{+20}_{-23}$ 
                                                 & 200 & $261^{+109}_{-96}$ \\
  &                                              & 400 & $400^{+29}_{-24}$ 
                                                 & 400 & $295^{+115}_{-108}$ \\
\midrule
\multirow{12}{*}{EMRI eccentricity} 
  & \multirow{4}{*}{$1\times100$~ks} & 0.01 & $0.0151^{+0.0026}_{-0.0045}$
                                     & 0.01 & $0.031^{+0.03}_{-0.01}$ \\
  &                                  & 0.1 & $0.105^{+0.011}_{-0.021}$ 
                                     & 0.1 & $0.097^{+0.03}_{-0.03}$ \\
  &                                  & 0.3 & $0.304^{+0.024}_{-0.027}$
                                     & 0.3 & $0.278^{+0.04}_{-0.04}$ \\
  &                                  & 0.5 & $0.526^{+0.020}_{-0.032}$
                                     & 0.5 & $0.474^{+0.07}_{-0.06}$ \\
\cmidrule(lr){2-6}
  & \multirow{4}{*}{$4\times100$~ks (200~ks gaps)} & 0.01 & $0.0104^{+0.0044}_{-0.0020}$
                                                   & 0.01 & $0.033^{+0.03}_{-0.01}$ \\
  &                                                & 0.1 & $0.128^{+0.014}_{-0.017}$ 
                                                   & 0.1 & $0.167^{+0.05}_{-0.06}$ \\
  &                                                & 0.3 & $0.319^{+0.092}_{-0.049}$ 
                                                   & 0.3 & $0.302^{+0.06}_{-0.05}$ \\
  &                                                & 0.5 & $0.521^{+0.040}_{-0.043}$ 
                                                   & 0.5 & $0.512^{+0.10}_{-0.08}$ \\
\cmidrule(lr){2-6}
  & \multirow{4}{*}{$4\times100$~ks (1~Ms gaps)} & 0.01 & $0.0323^{+0.016}_{-0.011}$ 
                                                 & 0.01 & $0.080^{+0.08}_{-0.04}$ \\
  &                                              & 0.1 & $0.0989^{+0.091}_{-0.017}$ 
                                                 & 0.1 & $0.162^{+0.09}_{-0.07}$ \\
  &                                              & 0.3 & $0.292^{+0.11}_{-0.028}$ 
                                                 & 0.3 & $0.294^{+0.11}_{-0.08}$ \\
  &                                              & 0.5 & $0.44^{+0.084}_{-0.078}$
                                                 & 0.5 & $0.461^{+0.14}_{-0.10}$ \\
\midrule
\multirow{3}{*}{EMRI inclination (degrees)} 
  & \multirow{1}{*}{$1\times100$~ks} & 60 & $58.9^{+20.2}_{-18.5}$ 
                                     & 60 & $57.480^{+19.81}_{-24.90}$ \\
\cmidrule(lr){2-6}
  & \multirow{1}{*}{$4\times100$~ks (200~ks gaps)} & 60 & $59.6^{+20.1}_{-19.3}$ 
                                                   & 60 & $56.280^{+20.24}_{-25.05}$ \\
\cmidrule(lr){2-6}
  & \multirow{1}{*}{$4\times100$~ks (1~Ms gaps)} & 60 & $60.3^{+19.0}_{-19.3}$ 
                                                 & 60 & $52.845^{+20.73}_{-23.84}$ \\
\midrule
\multirow{6}{*}{MBH spin} 
  & \multirow{2}{*}{$1\times100$~ks} & 0.5 & $0.478^{+0.33}_{-0.32}$ 
                                     & 0.5 & $0.505^{+0.31}_{-0.31}$ \\
  &                                  & 0.9 & $0.512^{+0.31}_{-0.34}$ 
                                     & 0.9 & $0.507^{+0.32}_{-0.31}$ \\
\cmidrule(lr){2-6}
  & \multirow{2}{*}{$4\times100$~ks (200~ks gaps)} & 0.5 & $0.471^{+0.31}_{-0.31}$ 
                                                   & 0.5 & $0.496^{+0.31}_{-0.31}$ \\
  &                                                & 0.9 & $0.497^{+0.30}_{-0.32}$ 
                                                   & 0.9 & $0.492^{+0.32}_{-0.31}$ \\
\cmidrule(lr){2-6}
  & \multirow{2}{*}{$4\times100$~ks (1~Ms gaps)} & 0.5 & $0.462^{+0.31}_{0.34}$ 
                                                 & 0.5 & $0.501^{+0.30}_{-0.30}$ \\
  &                                              & 0.9 & $0.43^{+0.33}_{-0.28}$ 
                                                 & 0.9 & $0.486^{+0.30}_{-0.28}$ \\
\midrule
\multirow{9}{*}{log(MBH mass/$M_\odot$)} 
  & \multirow{3}{*}{$1\times100$~ks} & 5 & $4.96^{+0.075}_{-0.058}$ 
                                     & 5 & $4.870^{+0.29}_{-0.20}$ \\
  &                                  & 6 & $5.93^{+0.14}_{-0.086}$ 
                                     & 6 & $5.680^{+0.26}_{-0.23}$ \\
  &                                  & 7 & ---
                                     & 7 & --- \\
\cmidrule(lr){2-6}
  & \multirow{3}{*}{$4\times100$~ks (200~ks gaps)} & 5 & $4.97^{+0.065}_{-0.055}$ 
                                                   & 5 & $4.885^{+0.30}_{-0.23}$ \\
  &                                                & 6 & $5.97^{+0.064}_{-0.055}$ 
                                                   & 6 & $5.880^{+0.29}_{-0.23}$ \\
  &                                                & 7 & $6.86^{+0.22}_{-0.14}$ 
                                                   & 7 & $6.060^{+0.31}_{-0.22}$ \\
\cmidrule(lr){2-6}
  & \multirow{3}{*}{$4\times100$~ks (1~Ms gaps)} & 5 & $4.97^{+0.07}_{-0.059}$ 
                                                 & 5 & $4.881^{+0.34}_{-0.24}$ \\
  &                                              & 6 & $5.92^{+0.12}_{-0.10}$ 
                                                 & 6 & $5.818^{+0.36}_{-0.30}$ \\
  &                                              & 7 & $6.92^{+0.17}_{-0.13}$ 
                                                 & 7 & $6.332^{+0.33}_{-0.29}$ \\
\midrule
\multirow{6}{*}{Disk inclination (degrees)} 
  & \multirow{2}{*}{$1\times100$~ks} & 5 & --- 
                                     & 5 & $8.698^{+5.52}_{-3.61}$ \\
  &                                  & 20 & --- 
                                     & 20 & $21.716^{+6.33}_{-7.27}$ \\
\cmidrule(lr){2-6}
  & \multirow{2}{*}{$4\times100$~ks (200~ks gaps)} & 5 & --- 
                                                   & 5 & $8.232^{+3.49}_{-3.99}$ \\
  &                                                & 20 & --- 
                                                   & 20 & $20.720^{+6.68}_{-7.93}$ \\
\cmidrule(lr){2-6}
  & \multirow{2}{*}{$4\times100$~ks (1~Ms gaps)} & 5 & --- 
                                                 & 5 & $13.108^{+7.81}_{-5.95}$ \\
  &                                              & 20 & --- 
                                                 & 20 & $22.822^{+8.57}_{-8.86}$\\
\midrule
\multirow{9}{*}{Disk precession period ($T_{\rm disk}/P_{\rm orb}$)} 
  & \multirow{3}{*}{$1\times100$~ks} & 5 & --- 
                                     & 5 & $46.659^{+23.79}_{-19.31}$ \\
  &                                  & 20 & --- 
                                     & 20 & $46.053^{+26.64}_{-18.70}$ \\
  &                                  & 50 & ---
                                     & 50 & $63.156^{+23.17}_{-23.38}$ \\
\cmidrule(lr){2-6}
  & \multirow{3}{*}{$4\times100$~ks (200~ks gaps)} & 5 & --- 
                                                   & 5 & $46.190^{+27.24}_{-20.93}$ \\
  &                                                & 20 & --- 
                                                   & 20 & $45.028^{+27.17}_{-20.32}$ \\
  &                                                & 50 & --- 
                                                   & 50 & $56.790^{+23.39}_{-19.85}$ \\
\cmidrule(lr){2-6}
  & \multirow{3}{*}{$4\times100$~ks (1~Ms gaps)} & 5 & --- 
                                                 & 5 & $47.691^{+26.43}_{-21.52}$ \\
  &                                              & 20 & --- 
                                                 & 20 & $45.416^{+25.79}_{-20.14}$ \\
  &                                              & 50 & --- 
                                                 & 50 & $57.260^{+23.83}_{-23.39}$ \\
\midrule
\bottomrule
\end{longtable*}

\end{document}